\documentclass[12pt]{article}
\usepackage{amsfonts,amsmath,amssymb,epsf,physics,braket,graphicx,hyperref,color,xcolor,comment,bm,tikz}
\usepackage{fancyhdr}
\usepackage{authblk}
\hypersetup{
    unicode=false,          
    pdftoolbar=true,        
    pdfmenubar=true,        
    linktocpage=true,       
    pdffitwindow=false,     
    pdfstartview={FitH},    
    pdfnewwindow=true,      
    colorlinks=true,       
    linkcolor=blue,          
    citecolor=blue,        
    filecolor=blue,      
    urlcolor=blue           
}
\usepackage[top=25truemm,bottom=28truemm,left=20truemm,right=20truemm]{geometry}
\numberwithin{equation}{section}

\title{\bf {\fontsize{30pt}{30pt}\selectfont Celestial Holography meets dS/CFT}}
\author[1]{
	Hideo~Furugori\thanks{\tt h-furugori(at)gauge.scphys.kyoto-u.ac.jp}
}
\author[2]{
	Naoki~Ogawa\thanks{
	\tt naoki.ogawa(at)yukawa.kyoto-u.ac.jp}
}
\author[1, 3]{
	Sotaro~Sugishita\thanks{
	\tt sotaro(at)gauge.scphys.kyoto-u.ac.jp}
	\vspace{5mm}
}
\author[2]{
	Takahiro~Waki\thanks{
	\tt takahiro.waki(at)yukawa.kyoto-u.ac.jp}
}
\affil[1]{\it\normalsize Department of Physics, Kyoto University, Kyoto 606-8502, Japan}
\affil[2]{\it\normalsize Center for Gravitational Physics and Quantum Information,\protect\\
Yukawa Institute for Theoretical Physics, Kyoto University, \protect\\
Kitashirakawa Oiwakecho, Sakyo-ku, Kyoto 606-8502, Japan}
\affil[3]{\it\normalsize RIKEN iTHEMS, Wako, Saitama 351-0198}
\date{}

\usetikzlibrary{calc}

\newcommand{\be}{\begin{equation}}           
\newcommand{\ee}{\end{equation}}             
\newcommand{\nn}{\nonumber \\}  
\newcommand{\dS}{\text{dS}}
\newcommand{\EAdS}{\text{EAdS}}

\numberwithin{equation}{section}  

\usepackage{amsthm}

\begin{document}
\maketitle
\thispagestyle{fancy}
\renewcommand{\headrulewidth}{0pt}

\begin{abstract}
We provide a concrete link between celestial amplitudes and cosmological correlators. 
We first construct a map from quantum field theories (QFTs) in $(D+2)$-dimensional Euclidean space $\mathbb{R}^{D+2}$ to theories on $(D+1)$-dimensional sphere, through a Weyl rescaling and a Fourier transformation. 
An analytic continuation extends this map to a relation between QFTs in Minkowski spacetime $\mathrm{M}_{D+2}$ and in de Sitter spacetime $\mathrm{dS}_{D+1}$ with the Bunch–Davies vacuum. 
Combining this relation with celestial holography, we show that the extrapolated operators in de Sitter space can be represented by operators on the celestial sphere $\mathrm{S}^{D}$.
Our framework offers a systematic route to transfer computational techniques and physical insights between celestial holography and the dS/CFT correspondence.

\end{abstract}

\newpage
\thispagestyle{empty}
\setcounter{tocdepth}{2}

\setlength{\abovedisplayskip}{12pt}
\setlength{\belowdisplayskip}{12pt}

\tableofcontents
\newpage

\section{Introduction}
The holographic principle in quantum gravity
 \cite{tHooft:1993dmi, Susskind:1994vu} asserts the following: 
a quantum gravity theory defined in a given spacetime is dual to a non‑gravitational theory that lives on a lower‑dimensional space.  
The well–known realization is the AdS/CFT correspondence \cite{Maldacena:1997re, Gubser:1998bc, Witten:1998qj}, and active studies have been extending this framework to the dS/CFT correspondence \cite{Witten:2001kn, Strominger:2001pn, Maldacena:2002vr}.  
More recently, a duality for asymptotically flat spacetimes has also been investigated, namely celestial holography \cite{Kapec:2016jld, Pasterski:2016qvg, Strominger:2017zoo, Pasterski:2017kqt, Pasterski:2017ylz, Donnay:2020guq, Ogawa:2022fhy}.

Celestial holography conjectures that scattering amplitudes in $(D+2)$-dimensional Minkowski spacetime can be described as correlation functions on
$D$-dimensional sphere.
More concretely, operators of the celestial CFT are obtained by, e.g., for a massless scalar, applying a Mellin transform \cite{Pasterski:2016qvg} as
\begin{align}
  \mathcal{O}_{\Delta}(k)\;=\;\int_{0}^{\infty}d\omega\,
  \omega^{\Delta-1}a(\omega k),
\end{align}
where $a$ is the annihilation operator of the massless scalar.
We can regard this $\mathcal{O}_{\Delta}$ as an operator that lives in the celestial sphere $\mathrm{S}^D$. 

In this paper, we provide a concrete link between the celestial amplitudes (i.e., the correlation functions of above $\mathcal{O}^{\pm}_{\Delta}$) and cosmological correlators in $(D+1)$-dimensional de Sitter space. 
Here, we mean by the cosmological correlators the asymptotic limit of the bulk correlators.  
More concretely, we consider the late or early limit of a bulk scalar field $\psi$ in dS in the global patch as
\begin{align}
\psi \sim e^{\mp\Delta_- t}O_{\Delta_-}^{\pm}+e^{\mp\Delta_+ t}O_{\Delta_+}^{\pm}
\qquad (t\rightarrow \pm\infty),
\end{align}
where $\Delta_\pm =D/2 \pm i \lambda$ with $\lambda \in \mathbb{R}$.
We call the correlation functions of $O_{\Delta_\pm}^{\pm}$ cosmological correlators. The extrapolated operators $O_{\Delta}^{\pm}$ are important objects in dS/CFT \cite{Strominger:2001pn, Bousso:2001mw}.
These $O_{\Delta_\pm}^{\pm}$ can also be considered as operators on $\mathrm{S}^D$ like the celestial operators $\mathcal{O}_{\Delta}$.

\begin{figure}
    \centering
    \includegraphics[width=0.8\linewidth]{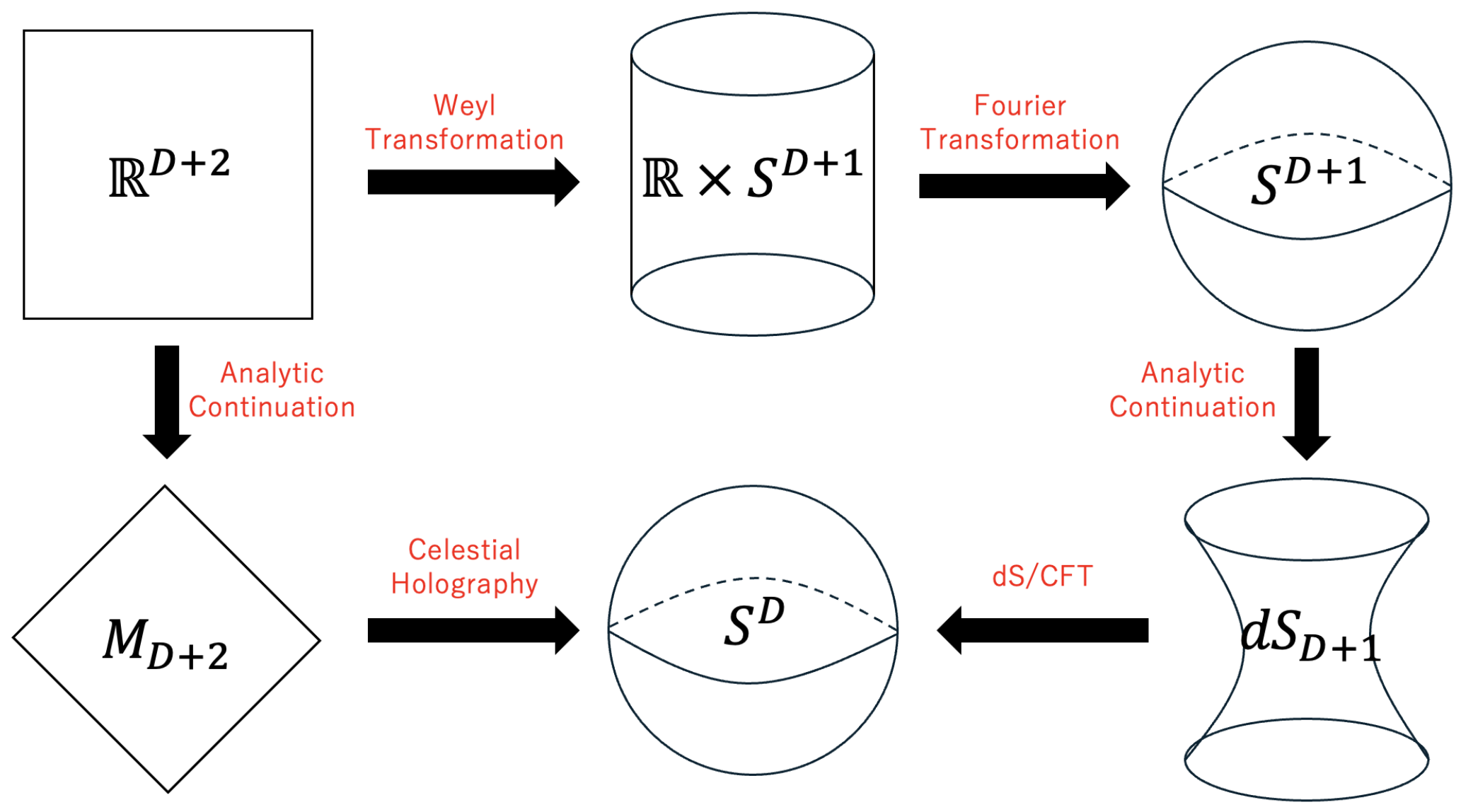}
    \caption{A schematic diagram illustrating how celestial holography meets the dS/CFT correspondence. }
    \label{fig:enter-label}
\end{figure}

We indeed obtain an identification rule between $\mathcal{O}_{\Delta}$ and $O_{\Delta}^{\pm}$. 
The strategy is as follows. 
We begin with a massless scalar field in flat Euclidean space
$\mathbb{R}^{D+2}$.  
A Weyl transformation maps the theory to the cylindrical space
$\mathrm{S}^{D+1}\times\mathbb{R}$.  
Upon Fourier transformation along the $\mathbb{R}$ direction, we obtain a theory on the sphere $\mathrm{S}^{D+1}$. This procedure is given in section~\ref{sec2}.
We then analytically continue this theory on $\mathrm{S}^{D+1}$ to a theory in dS$_{D+1}$ with the Bunch–Davies vacuum \cite{Bunch:1978yq, Mottola:1984ar, Allen:1985ux, Marolf:2010zp, Marolf:2010nz, Higuchi:2010xt}. Under this analytic continuation, the original $\mathbb{R}^{D+2}$ theory becomes a theory in Minkowski space $\mathrm{M}_{D+2}$.  
We thus obtain the relation between correlation functions in dS$_{D+1}$ and $\mathrm{M}_{D+2}$ in section~\ref{sec3}. 
In section~\ref{sec4}, we apply a dictionary of celestial holography \cite{Furugori:2023hgv}, and we find that the extrapolated operators in  $O_{\Delta}^{\pm}$ can be written as a linear combination of the celestial operators $\mathcal{O}_{\Delta}$ and the shadow operators (the definition of the shadow operators will be given below). 
Our procedures are schematically summarized in Figure~\ref{fig:enter-label}.



Our work is largely motivated by \cite{Cheung:2016iub, deBoer:2003vf} (see also recent related works \cite{Ogawa:2022fhy, Ball:2019atb,  Iacobacci:2022yjo, Sleight:2023ojm, Iacobacci:2024nhw,Jorstad:2023ajr}), which propose that $\mathrm{M}_{D+2}$ can be regarded, via a kind of ``dimensional reduction''\footnote{This is not the standard dimensional reduction because we consider a mode expansion on a non-compact space (i.e., Fourier transformations).}, as
\begin{align}
  \mathrm{M}_{D+2}\;\sim\;
  \mathrm{dS}_{D+1}\;+\;\mathrm{EAdS}_{D+1}^{+}\;+\;\mathrm{EAdS}_{D+1}^{-}.
\end{align}
In our framework, we analytically continue a theory on a sphere.
This procedure makes it clearer why the Bunch-Davies vacuum (Euclidean vacuum) appears, because it is well known that the dS correlation functions are obtained by the analytic continuation from the Euclidean theory on the sphere. 
We will confirm a connection with earlier results \cite{Cheung:2016iub, deBoer:2003vf} by deforming the contours of the path-integral appropriately in section~\ref{sec3}.
In the same spirit as these previous works, we expect that our identification provides us with a chance to import rich results from (A)dS/CFT into celestial holography, e.g. evaluation of the central charge in $\text{(A)dS}_{3}/\text{CFT}_{2}$ \cite{Strominger:2001pn, Strominger:1997eq, Brown:1986nw}.

We further note that Ref.~\cite{Sleight:2023ojm} proposes a new celestial dictionary motivated by the extrapolate dictionary in (A)dS/CFT, and  \cite{Jorstad:2023ajr} investigates a relation between this new dictionary and the conventional celestial dictionary \cite{Pasterski:2016qvg}. 
Ref.~\cite{Pasterski:2016qvg} also obtained a relation between the celestial operators $\mathcal{O}_\Delta$ and extrapolated operators in (A)dS. Although their method and results are similar to ours, there are slight differences. 
We will argue the reasons for the differences later.

\subsubsection*{Notation and useful formulas}
Before proceeding, we summarize our notation and several useful identities here.

We define the measure \(D^D p\) of conformal integrals followed by \cite{Simmons-Duffin:2012juh} 
\begin{align}
    \int D^D f(p)
    &\equiv \frac{2}{\mathrm{Vol}\,\mathrm{GL}(1,\mathbb{R})^+}
      \int_{p^0>0} d^{D+2}p\;\delta\left(p^2\right)f(p),
\end{align}
where $f(p)$ should have a degree $-D$ under the rescaling $p \to a p$.
If we take a ``gauge'' \(p=(1,\bm{p})\) where \(\bm{p}\) is a unit vector in \(\mathbb{R}^{D+1}\), we have  
\begin{align}
    \int D^D p\,f(p)
    =\int_{\mathrm{S}^D} d^D\bm{p}\;f(p)| _{p=(1,\bm{p})}.
\end{align}

When integrating over a manifold \(M\), we write:
\begin{align}
    \int_M[d^Dx]\equiv\int_Md^Dx\sqrt{g_M}
\end{align}
where $g_M$ is the determinant of the metric of the manifold $M$.

In the celestial holography, where we adopt the rectified dictionary \cite{Furugori:2023hgv}\footnote{``Rectified'' means that we take the shadow transformations for the Mellin transformation of the creation operators $a^\dagger$. This rectification for massless bulk fields is required to maintain consistency with the massless limit of the celestial dictionary for massive bulk fields as shown in \cite{Furugori:2023hgv}.}, it is supposed that the creation and annihilation operators of bulk fields on the flat space are related to primary and their shadow operators in the celestial CFT via the Mellin transformation as
\begin{subequations}\label{celestial_dic}
\begin{align}
  \mathcal{O}_{\Delta}^{+}(p)
    &= \int_0^\infty d\omega\;\omega^{\Delta-1}a_{+}(\omega \bm{p}), 
    &\widetilde{\mathcal{O}_{D-\Delta}^{\prime +}}(p)
    &= \int_0^\infty d\omega\;\omega^{\Delta-1}a^\dagger_{+}(\omega \bm{p}),\\
  \widetilde{\mathcal{O}_{D-\Delta}^{-}}(p)
    &= \int_0^\infty d\omega\;\omega^{\Delta-1}a^\dagger_{-}(\omega \bm{p}),
    &\mathcal{O}_{\Delta}^{\prime-}(p)
    &= \int_0^\infty d\omega\;\omega^{\Delta-1}a_{-}(\omega \bm{p}),
\end{align}
\end{subequations}
where the label `$+$/$-$' on the creation and annihilation operators represents the outgoing/incoming modes.\footnote{\label{footO'}We define $\mathcal{O}'$ as the shadow transform of the Mellin transform of $a^\dagger$, so that every two-point function of non-shadow operators takes a power-law form. 
Another reason that we take the shadow for $\widetilde{\mathcal{O}_{D-\Delta}^{\prime +}}$ is that we should have $\widetilde{\mathcal{O}_{D-\Delta}^{\prime +}}(\bm{p})=\widetilde{\mathcal{O}_{D-\Delta}^{-}}(-\bm{p})$ for the free limit  because $a_+^\dagger=a_-^\dagger$ for the free theory. 
In addition, the overall normalization of $\mathcal{O}$ and $\mathcal{O}^\prime$ is rescaled compared to \cite{Furugori:2023hgv} for simplicity.
}
The shadow transformation $\widetilde{\mathcal{O}_\Delta}$ is defined by 
\begin{align}\label{def:shadow}
    \widetilde{\mathcal{O}_{\Delta}}(p)
  = \frac{1}{S_{\Delta}}
    \int d^D k\;(-2p\!\cdot\!k)^{\Delta-D}
    \mathcal{O}_{\Delta}(k).
\end{align}
The normalization factor $S_{\Delta} $ is given by
\begin{align}
  S_{\Delta}
  = 2^{2\Delta-D}\pi^{h}
    \frac{\Gamma(\Delta-h)}{\Gamma(D-\Delta)}, 
  \qquad
  h = \frac{D}{2},
\end{align}
which is chosen such that the transform is involutive\footnote{The involutive condition \eqref{involutive} does not uniquely fix $S_{\Delta}$. For instance, we may choose  $S_{\Delta} = \pi^{h}
    \frac{\Gamma(\Delta-h)}{\Gamma(D-\Delta)}$. We add the factor $2^{2\Delta-D}$ for later convenience.}:
\begin{align}
  \label{involutive}
  \widetilde{\left(\widetilde{\mathcal{O}_{\Delta}}\right)}
  = \mathcal{O}_{\Delta}.
\end{align}

Finally, the following integral identities are useful for the shadow transformations \cite{Simmons-Duffin:2012juh}:
\begin{align}
\label{shadow-formula1}
  \int D^D k\;
  \frac{1}{\left(-2p\!\cdot\!k\right)^{D-\Delta}}
  \frac{1}{\left(-2X\!\cdot\!k\right)^{\Delta}}
  &=
  \pi^{h}
  \frac{\Gamma(\Delta-h)}{\Gamma(\Delta)}
  \frac{(-X^2)^{ h-\Delta}}
       {\left(-2 p\!\cdot\!X\right)^{D-\Delta}},\\
  \int D^D k\;
  \frac{1}{\left(-2 p\!\cdot\!k\right)^{h+i\lambda}}
  \frac{1}{\left(-2 q\!\cdot\!k\right)^{h-i\lambda}}
  &=
  \pi^{D}
  \frac{\Gamma(-i\lambda) \Gamma(i\lambda)}
       {\Gamma(h-i\lambda) \Gamma(h+i\lambda)}
   \delta^{(D)}(p-q),
   \label{shadow-formula2}
\end{align}
where $p$ and $q$ are null vectors and $X$ is an arbitrary non-null vector, and $\delta^{(D)}(p-q)$ represnts a delta function for the measure $D^D p$.

We will relate these celestial operators $\mathcal{O}_\Delta$ to extrapolated operators $O_\Delta$ in the late/early time limit in dS. 
The notation in this paper is summarized in Table~\ref{tab:notation_holograpy}.
\begin{table}
    \centering
    \begin{tabular}{c|c|c}
         & $\mathrm{M}_{D+2}$ & $\mathrm{dS}_{D+1}$ \\
        Bulk operators & $\Phi$ & $\psi_\lambda$ \\
       Operators on  $\mathrm{S}^{D}$  & $\mathcal{O}_\Delta$ & $O_\Delta$
    \end{tabular}
    \caption{Our notation of operators}
    \label{tab:notation_holograpy}
\end{table}

$\mathcal{O}_\Delta$ and $\mathcal{O}'_\Delta$ are not independent operators because they are related by the Hermitian conjugation.\footnote{Here the Hermitian conjugation $\dagger$ is the conjugation with respect to the inner product on the Minkowski spacetime, which does not imply the conjugation in the celestial CFT. See also \cite{Crawley:2021ivb}.
A similar relation between the bulk Hermitian conjugation and the shadow transformation is also observed in dS \cite{SalehiVaziri:2024joi}.
}
Indeed, taking $\Delta_\pm =D/2 \pm i \lambda$ with $\lambda \in \mathbb{R}$, we have 
\begin{align}
\label{eq:Odagger}
   \left( \mathcal{O}_{\Delta_\pm}^{+} \right)^\dagger = \widetilde{\mathcal{O}_{\Delta_\pm }^{\prime +}}, \qquad  \left( \mathcal{O}_{\Delta_\pm}^{-} \right)^\dagger = \widetilde{\mathcal{O}_{\Delta_\pm }^{\prime -}}.
\end{align}

\section{QFT on $\mathrm{S}^{D+1}$ from QFT on $\mathbb{R}^{D+2}$}\label{sec2}
In this section, we construct a map from QFT on $\mathbb{R}^{D+2}$ to QFT on $\mathrm{S}^{D+1}$ via the Weyl and Fourier transformations. 
It can be regarded as the Euclidean version of the construction in \cite{Cheung:2016iub,deBoer:2003vf}.

\subsection{Free Field}
We begin by considering a free real scalar field on the Euclidean space \(\mathbb{R}^{D+2}\) as
\begin{align}
    S \;=\;\int_{\mathbb{R}^{D+2}} [d^{D+2}X]\left(\frac12 G^{IJ}\nabla_I\Phi \nabla_J\Phi 
+\frac{D}{8(D+1)} R \Phi^2\right).
\end{align}
Here, the metric is given by 
\begin{align}
ds^2=G^{IJ}dX_IdX_J =(dX^1)^2 + \cdots + (dX^{D+2})^2.
\end{align}
We can rewrite the metric by using polar coordinates,
\begin{align}
    ds^2=dr^2+r^2 d\Omega^2_{D+1} 
\;=\;e^{2\xi}\left(d\xi^2 + d\Omega^2_{D+1} \right),
\end{align}
where $d\Omega^2_{D+1}$ denotes the line elements on \(\mathrm{S}^{D+1}\) and $r=e^\xi$.

Next, we perform a Weyl transformation and redefine the coordinates and the field as follows:
\begin{align}
    g_{IJ} \;&=\; e^{-2\xi}G_{IJ},\\
    \phi \;&=\; e^{\frac{D}{2}\xi}\Phi.
\end{align}
Under this transformation, the theory originally defined in \(\mathbb{R}^{D+2}\) is mapped to \(\mathrm{S}^{D+1}\times \mathbb{R}\), and the action becomes
\begin{align}
S \;=\;\int_{S^{D+1}} [d^{D+1}x]\int d\xi\;\left(
\frac12 g_S^{ij}\nabla_i\phi \nabla_j\phi \;+\;\frac12 \partial_\xi\phi \partial_\xi\phi 
\;+\;\frac{D^2}{8} \phi^2
\right),
\end{align}
where $g_S$ is the metric on \(\mathrm{S}^{D+1}\).

Furthermore, we implement the Fourier transformation along the \(\xi\) direction:
\begin{align}
    \begin{aligned}
        \psi_\lambda(x) \;&=\;\frac{1}{\sqrt{2\pi}}\int^\infty_{-\infty}\!\! d\xi\;e^{-i\lambda\xi} \phi(\xi,x),\\
        \phi(\xi,x) \;&=\;\frac{1}{\sqrt{2\pi}}\int^\infty_{-\infty}\!\! d\lambda\;e^{i\lambda\xi} \psi_\lambda(x), 
        \label{mode_exp}
    \end{aligned}
\end{align}
This transformation leads to an action for an infinite number of fields $\psi_\lambda$ on \(\mathrm{S}^{D+1}\): 
\begin{align}
\label{LofS}
S \;=\;\int_{\mathrm{S}^{D+1}} [d^{D+1}x]\int_{0}^{\infty}d\lambda\left(
g_S^{ij}\nabla_i\psi_{\lambda}^\dagger\nabla_j\psi_{\lambda}
+\left(\frac{D^2}{4}+\lambda^2\right) \psi_\lambda^\dagger\psi_{\lambda}
\right),
\end{align}
where the range of $\lambda$ is restricted to $(0,\infty)$ in the action. 
Note that we have $\psi_{\lambda}^\dagger=\psi_{-\lambda}$ from the definition \eqref{mode_exp} because $\phi$ is a real field. 
Due to the Fourier transformation performed on a noncompact manifold, we have a continuous label $\lambda$. 

In summary, the original field \(\Phi\) in \(\mathbb{R}^{D+2}\) and the field \(\psi_\lambda\) on \(\mathrm{S}^{D+1}\) are related by 
\begin{align}
\label{eq:psi-Phi}
\psi_\lambda(x) \;=\;\frac{1}{\sqrt{2\pi}}\int d\xi \, e^{\Delta_-\xi}{e^{-\epsilon r}} \Phi\left(X(\xi,x)\right)
\;=\;\frac{1}{\sqrt{2\pi}}\int \frac{dr}{r}\;r^{ \Delta_-}{e^{-\epsilon r}}\Phi\left(X(r,x)\right)
\end{align}
with 
\begin{align}
    \Delta_\pm={D}/{2}\pm i\lambda,
\end{align}
where we have introduced an IR cut-off via $e^{-\epsilon r}$.  
Each mode \(\psi_\lambda\) has a mass
\begin{align}
\label{m_lambda}
m_\lambda = \sqrt{\frac{D^2}{4} + \lambda^2}.
\end{align}

An explicit check of the equivalence of the two-point functions between the sphere and the plane is given in Appendix~\ref{2dsphere}.

\subsection{Interactions}
We consider the \(\Phi^k\) interaction on $\mathbb{R}^{D+2}$ which is given by
\begin{align}
\label{S-int}
S_\mathrm{int}=\int d^{D+2}X \sqrt{G} \frac{\alpha_k}{k!} \Phi^k,
\end{align}
where $\alpha_k$ is a coupling constant.
After the Weyl transformation, this term becomes
\begin{align}
S_\mathrm{int}=\int [d^{D+1}x] \int d \xi \,\frac{\alpha_k}{k!} \exp{\left(D+2-\frac{kD}{2}\right)\xi} \phi^k.
\end{align}
Using the mode expansion \eqref{mode_exp}, the interaction takes the form
\begin{align}
\label{L_int}
S_\mathrm{int}=\frac{\alpha_k}{k!}\int d^{D+1}x \sqrt{g} \int\prod_{i=1}^k \frac{d\lambda_i}{\sqrt{2\pi}}\, 2\pi\delta^G\left(-\sum_{i=1}^k \lambda_i+i(D+2-kD/2)\right) \prod_{i=1}^k \psi_{\lambda_i},
\end{align}
where $\psi_{-\lambda}\equiv\psi^\dagger_\lambda$, and $\delta^G(z)$ is the generalized delta function defined by \(\delta^G(z)=\frac{1}{2\pi}\int_{0}^\infty  \frac{dr}{r} r^{-iz}\) where we need an appropriate regularization \cite{Donnay:2020guq}. For real $z$, it reduces to the ordinary delta function. Thus, in the Weyl invariant interaction case, which corresponds to \( k = 2(D+2)/{D} \), the generalized delta function reduces to the ordinary delta function.
We will consider a $k=4$ interaction for $D=2$ as an example later.

\section{QFT in dS$_{D+1}$ from QFT in {M}\(_{D+2}\)  }\label{sec3}

From the correspondence discussed in the previous section, we have the following formula relating the \(n\)-point functions on the sphere \(\mathrm{S}^{D+1}\) to ones on Euclidean space \(\mathbb{R}^{D+2}\): 
\begin{align}
\langle \psi_{\lambda_1}(x_1)\cdots \psi_{\lambda_n}(x_n)\rangle_{\mathrm{S}^{D+1}}
= \int \biggl[\prod_{i=1}^n\frac{1}{\sqrt{2\pi}} \frac{dr_i}{r_i} r_i^{\Delta_{i-}}e^{-\epsilon r_i}\biggr] \langle \Phi(r_1,x_1)\cdots \Phi(r_n,x_n)\rangle_{\mathbb{R}^{D+2}}.
\label{psi-Phi3}
\end{align}
Introducing the polar coordinate as\footnote{We choose $x^{D+2}=-\cos\theta$ instead of $x^{D+2}=+\cos\theta$ so that the direction of increasing $\theta$ is aligned with the $x^{D+2}$-axis. With this choice, the analytic continuation of Eq.~\eqref{theta_i-t_i} matches the standard convention for continuing between Lorentzian and Euclidean time \eqref{ac:time}.} 
\begin{align}
x^{D+2}=-\cos\theta,
  \qquad
  \bm{x}=(x^1,\ldots ,x^{D+1}) = \sin\theta \,\hat n,
\end{align}
we analytically continue the formula \eqref{psi-Phi3} by continuing the polar angle \(\theta_i\) in the spherical coordinates of the external vertices as
\begin{align}
\theta_i = \frac{\pi}{2} + i t_i\,.
\label{theta_i-t_i}
\end{align}
Under this analytic continuation, the correlation functions in Euclidean space \(\mathbb{R}^{D+2}\) can be identified with those for the standard Poincaré vacuum of Minkowski spacetime \(\mathrm{M}_{D+2}\) (with external vertices restricted to the spherical Rindler region \(X^2>0\)), whereas the correlation functions on \(\mathrm{S}^{D+1}\) can be viewed as those in de Sitter spacetime \(\mathrm{dS}_{D+1}\) in the Bunch-Davies (BD) vacuum (the Euclidean vacuum):
\begin{align}
\langle \psi_{\lambda_1}(x_1)\cdots \psi_{\lambda_n}(x_n)\rangle_{\mathrm{dS}^{D+1}}
= \int \biggl[\prod_{i=1}^n\frac{1}{\sqrt{2\pi}} \frac{dr_i}{r_i} r_i^{\Delta_{i-}}e^{-\epsilon r_i}\biggr] \langle \Phi(r_1,x_1)\cdots \Phi(r_n,x_n)\rangle_{\mathrm{M}_{D+2}}.
\label{psi-Phi4}
\end{align}

\begin{figure}[htbp]
        \centering

        \begin{tikzpicture}[scale=0.9]
            \coordinate[label = right:$~$] (I)   at (pi, 0);
            \coordinate[label = above:$~$] (II)  at (0, pi);
            \coordinate[label = left:$~$]  (III) at (-pi, 0);
            \coordinate[label = below:$~$] (IV)  at (0, -pi);
    
            \draw (I) -- 
            node[midway, above right]    {$\mathcal{I}^+$}
            (II) -- 
            node[midway, above left]    {$\mathcal{I}^+$}
            (III) -- 
            node[midway, below left]    {$\mathcal{I}^-$}
            (IV) -- 
            node[midway, below right]    {$\mathcal{I}^-$} cycle;
    
            \coordinate (M_I_II)   at (pi / 2, pi / 2);
            \coordinate (M_II_III) at (-pi / 2, pi / 2);
            \coordinate (M_III_IV) at (-pi / 2, -pi / 2);
            \coordinate (M_IV_I)   at (pi / 2, -pi / 2);
    
            \draw[very thin] (M_I_II) -- (M_III_IV);
    
            \draw[very thin] (M_II_III) -- (M_IV_I);
    
            \draw[densely dotted, blue, ultra thick] plot[variable = \t, smooth] ({rad(atan(12 * cosh(\t)  / (1 - 36)))},{rad(atan(12 * sinh(\t) / (1 + 36)))});
            \draw[densely dotted, blue, ultra thick] plot[variable = \t, smooth] ({rad(atan(5.4 * cosh(\t)  / (1 - 7.29)))},{rad(atan(5.4 * sinh(\t) / (1 + 7.29)))});
            \draw[densely dotted, blue, ultra thick] plot[variable = \t, smooth] ({rad(atan(3 * cosh(\t)  / (1 - 2.25)))},{rad(atan(3 * sinh(\t) / (1 + 2.25)))});
            \draw[densely dotted, blue, ultra thick] (M_II_III) -- (M_III_IV);
            \draw[densely dotted, blue, ultra thick] plot[variable = \t, smooth] ({-pi + rad(atan((4 / 3) * cosh(\t)  / (1 - (4 / 9))))},{rad(atan((4 / 3) * sinh(\t) / (1 + (4 / 9))))});
            \draw[densely dotted, blue, ultra thick] plot[variable = \t, smooth] ({-pi + rad(atan((2 / 2.7) * cosh(\t)  / (1 - (1 / 7.29))))},{rad(atan((2 / 2.7) * sinh(\t) / (1 + (1 / 7.29))))});
            \draw[densely dotted, blue, ultra thick] plot[variable = \t, smooth] ({-pi + rad(atan((1 / 3) * cosh(\t)  / (1 - (1 / 36))))},{rad(atan((1 / 3) * sinh(\t) / (1 + (1 / 36))))});
    
            \draw[red,very thick] plot[variable = \t, smooth] ({rad(atan(12 * sinh(\t)  / (1 + 36)))},{pi + rad(atan(12 * cosh(\t) / (1 - 36)))}) ;
            \draw[red,very thick] plot[variable = \t, smooth] ({rad(atan(5.4 * sinh(\t)  / (1 + 7.29)))},{pi + rad(atan(5.4 * cosh(\t) / (1 - 7.29)))});
            \draw[red,very thick] plot[variable = \t, smooth] ({rad(atan(3 * sinh(\t)  / (1 + 2.25)))},{pi + rad(atan(3 * cosh(\t) / (1 - 2.25)))});
            \draw[red,very thick] (M_III_IV) -- (M_IV_I);
            \draw[red,very thick] plot[variable = \t, smooth] ({rad(atan((4 / 3) * sinh(\t)  / (1 + (4 / 9))))},{rad(atan((4 / 3) * cosh(\t) / (1 - (4 / 9))))});
            \draw[red,very thick] plot[variable = \t, smooth] ({rad(atan((2 / 2.7) * sinh(\t)  / (1 + (1 / 7.29))))},{rad(atan((2 / 2.7) * cosh(\t) / (1 - (1 / 7.29))))});
            \draw[red,very thick] plot[variable = \t, smooth] ({rad(atan((1 / 3) * sinh(\t)  / (1 + (1 / 36))))},{rad(atan((1 / 3) * cosh(\t) / (1 - (1 / 36))))});
    
            \draw[densely dotted, blue, ultra thick] plot[variable = \t, smooth] ({pi + rad(atan(12 * cosh(\t)  / (1 - 36)))},{rad(atan(12 * sinh(\t) / (1 + 36)))});
            \draw[densely dotted, blue, ultra thick] plot[variable = \t, smooth] ({pi + rad(atan(5.4 * cosh(\t)  / (1 - 7.29)))},{rad(atan(5.4 * sinh(\t) / (1 + 7.29)))});
            \draw[densely dotted, blue, ultra thick] plot[variable = \t, smooth] ({pi + rad(atan(3 * cosh(\t)  / (1 - 2.25)))},{rad(atan(3 * sinh(\t) / (1 + 2.25)))});
            \draw[densely dotted, blue, ultra thick] (M_IV_I) -- (M_I_II);
            \draw[densely dotted, blue, ultra thick] plot[variable = \t, smooth] ({rad(atan((4 / 3) * cosh(\t)  / (1 - (4 / 9))))},{rad(atan((4 / 3) * sinh(\t) / (1 + (4 / 9))))});
            \draw[densely dotted, blue, ultra thick] plot[variable = \t, smooth] ({rad(atan((2 / 2.7) * cosh(\t)  / (1 - (1 / 7.29))))},{rad(atan((2 / 2.7) * sinh(\t) / (1 + (1 / 7.29))))});
            \draw[densely dotted, blue, ultra thick] plot[variable = \t, smooth] ({rad(atan((1 / 3) * cosh(\t)  / (1 - (1 / 36))))},{rad(atan((1 / 3) * sinh(\t) / (1 + (1 / 36))))});
    
            \draw[red,very thick] plot[variable = \t, smooth] ({rad(atan(12 * sinh(\t)  / (1 + 36)))},{rad(atan(12 * cosh(\t) / (1 - 36)))}) ;
            \draw[red,very thick] plot[variable = \t, smooth] ({rad(atan(5.4 * sinh(\t)  / (1 + 7.29)))},{rad(atan(5.4 * cosh(\t) / (1 - 7.29)))});
            \draw[red,very thick] plot[variable = \t, smooth] ({rad(atan(3 * sinh(\t)  / (1 + 2.25)))},{rad(atan(3 * cosh(\t) / (1 - 2.25)))});
            \draw[red,very thick] (M_I_II) -- (M_II_III);
            \draw[red,very thick] plot[variable = \t, smooth] ({rad(atan((4 / 3) * sinh(\t)  / (1 + (4 / 9))))},{- pi + rad(atan((4 / 3) * cosh(\t) / (1 - (4 / 9))))});
            \draw[red,very thick] plot[variable = \t, smooth] ({rad(atan((2 / 2.7) * sinh(\t)  / (1 + (1 / 7.29))))},{- pi + rad(atan((2 / 2.7) * cosh(\t) / (1 - (1 / 7.29))))});
            \draw[red,very thick] plot[variable = \t, smooth] ({rad(atan((1 / 3) * sinh(\t)  / (1 + (1 / 36))))},{- pi + rad(atan((1 / 3) * cosh(\t) / (1 - (1 / 36))))});
        \end{tikzpicture}

        \caption{Penrose diagram of Minkowski space and the Milne slices. Red curves represent $\mathrm{EAdS}$ slices, and blue dotted curves represent $\mathrm{dS}$ slices. By performing the Fourier transformation, we obtain a picture of $\mathrm{EAdS}^++\mathrm{dS}+\mathrm{EAdS}^-$.}
        \label{Milne}
    \end{figure}
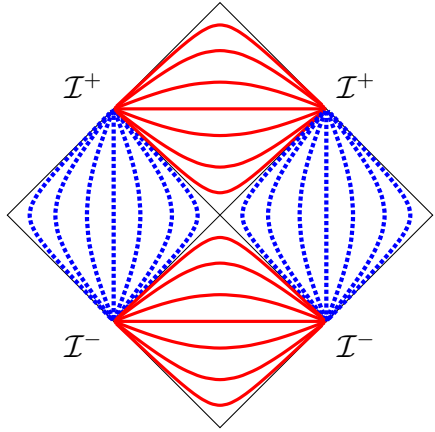

One might think that this relation between the BD vacuum in $\mathrm{dS}$ and the Poincaré vacuum in the Minkowski space is counterintuitive, because if one adopts a foliation of Minkowski space (see Fig.\ref{Milne}), one would typically obtain a theory defined on \(\mathrm{EAdS}^+ + \mathrm{EAdS}^- + \mathrm{dS}\) \cite{Cheung:2016iub, deBoer:2003vf}.

\begin{figure}[htbp]
  \centering
  \begin{tikzpicture}[scale=1.0]
    \draw[->] (-0.5,0) -- (7,0) node[right] {Re$(\theta)$};
    \node at (0,0) [above] {$0$};
    \node at (3,0) [above right] {$\pi/2$};
    \node at (6,0) [below] {$\pi$};

    \draw[red, thick, ->] (0,0) -- (0,-3) node[midway, left] {$A$};
    \node at (0,-3) [below] {$-i\infty$};
    \draw[densely dashed, ->] (0,-3) -- (3,-3);

    \draw[blue, thick, ->] (3,-3) -- (3,3) node[midway, below right] {$B$};
    \node at (3,-3) [below] {$\pi/2-i\infty$};
    \node at (3,3) [above] {$\pi/2+i\infty$};
    \draw[densely dashed, ->] (3,3) -- (6,3);

    \draw[red, thick, ->] (6,3) -- (6,0) node[midway, right] {$C$};
    \node at (6,3) [above] {$\pi+i\infty$};
  \end{tikzpicture}
  
 \caption{Deformation of the integral contour of the polar angle $\theta$ in the path-integral. The original contour is $0 \to \pi$ along the real axis. It is deformed into the contours $A, B, C$ (and the dotted lines, which can be ignored). }
  \label{fig:theta-contours}
\end{figure}
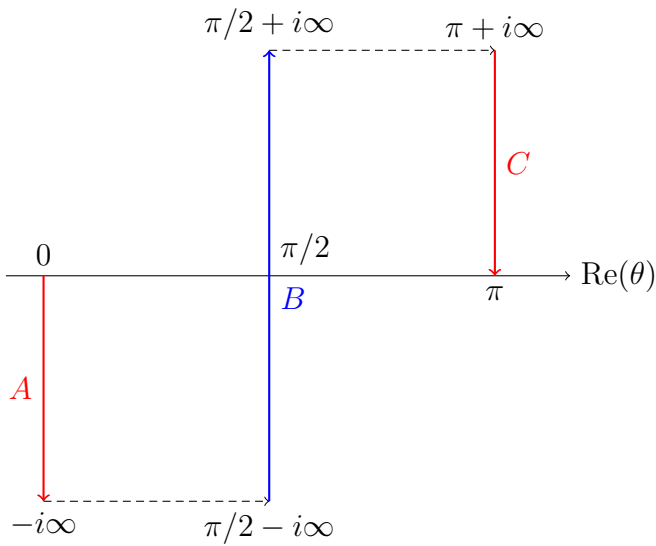

Nevertheless, our construction remains compatible with these earlier approaches. 
Specifically, we analytically continue the external vertices on \( \mathrm{S}^{D+1} \) and appropriately deform the integration contour for the polar angle $\theta\in [0,\pi]$ in the path-integral to the contour shown below (depicted in Fig.~\ref{fig:theta-contours}):\footnote{When deforming the contour into the paths \(A + B + C\), the real part of the coordinate along path must increase monotonically. Otherwise, the contour may cross the branch cut of the integrand, which is expressed as a product of propagators. For this reason, path cannot be exactly vertical; it must be slightly tilted.}
\begin{equation}
\begin{aligned}
A: &\quad \theta = -i\chi,        & \chi &: 0 \to \infty, \\
B: &\quad \theta = \frac{\pi}{2} + it, & t    &: -\infty \to \infty, \\
C: &\quad \theta = \pi - i\chi,   & \chi &: -\infty \to 0 .
\end{aligned}
\label{eq:theta-contours}
\end{equation}
We assume here that the integrand decays sufficiently rapidly at infinity so that we can perform this change of the contour.
Thus, the path-integral on the sphere can be deformed to the path-integral on  \(A + B +C\).

We can see that the path \(B\) corresponds to \(\mathrm{dS}\), \(A\) does to \(\mathrm{EAdS}^-\), and \(C\) does to \(\mathrm{EAdS}^+\),\footnote{The details of this statement in the Poincaré patch version can be found in \cite{Sleight:2021plv}.}
although we confine further details to Appendix~\ref{app:contour-split}, 
The $\mathrm{EAdS}$ parts play the role of preparing the Bunch-Davies vacuum for the initial and final states in dS.\footnote{In \cite{Maldacena:2002vr, Harlow:2011ke}, it is argued that the path-integral on Euclidean AdS produces the wave-fucntions for the Bunch-Davies vacuum for the initial states.} 
In this sense, our prescription is consistent with previous constructions.

While there is indeed no contradiction, interpreting this setup as a theory in \(\mathrm{EAdS}^+ + \mathrm{EAdS}^- + \mathrm{dS}\) requires a rather unconventional quantum field theory with the following features:
\begin{itemize}
\item \textbf{Physical interpretation of dS part:} It is crucial to emphasize that if we consider the path-integral only on the dS segment with the standard Feynman's $i\epsilon$ prescription, the resulting correlation functions do not match with those in the Bunch–Davies (BD) vacuum.\footnote{See \cite{Fukuma:2013mx} where such a two-point function is explicitly computed and it is different from the in-in two-point function with the Bunch-Davies vacuum.} 
In contrast, the full sum over the entire path, comprising
$\mathrm{EAdS}^+ + \mathrm{EAdS}^- + \mathrm{dS}$, does yield the correlation function in the BD vacuum of de Sitter space.
\item \textbf{Interactions across EAdS and dS:} If we consider interacting theories, the fields defined on EAdS and dS should interact with each other in the sense that the propagator connecting points in EAdS and dS can contribute to correlation functions. Since we usually do not consider interactions among fields on disconnected manifolds in the standard QFT, we might extend this standard framework.
\item \textbf{Propagators on EAdS:} Even for propagators with endpoints in EAdS, one must use analytically continued propagators originating from those on \(\mathrm{S}^{D+1}\). These propagators generally differ from the standard ones typically used in EAdS, which are defined with Dirichlet boundary conditions.
\end{itemize}
For the reasons outlined above, a direct interpretation in terms of theory in dS seems to be appropriate, particularly in the context of holographic applications, while the interpretation as theories on \(\mathrm{EAdS}^+ + \mathrm{EAdS}^- + \mathrm{dS}\) is interesting and also valuable as a computational tool.

\subsubsection*{Subtleties in the Analytic Continuation}
Finally, let us point out a few subtle issues that require attention. Since Eq.\eqref{psi-Phi4} is obtained by analytically continuing both sides of Eq.\eqref{psi-Phi3}, one must first perform the Mellin transformation with respect to the coordinates $r_i$ of the external vertices in the R.H.S. of \eqref{psi-Phi3}, and only afterwards carry out the analytic continuation $\theta = \pi/2 + it$ together with the deformation of the path-integral contour. Following this procedure, the Mellin transformation produce a factor such as conformal primary wave : $(-pr\cdot x + i\epsilon)^{-\Delta}$ (where $r$ denotes the coordinate of the internal vertex), which introduces a nontrivial branch cut in the $r$-plane. Since the contour deformation for the internal vertex must subsequently be performed, the following subtlety arise.

In such cases, if one attempts to compute the Minkowski correlator from the Euclidean correlator by following the standard prescription in QFT---namely, rotating integral contour of the internal vertex as \(X^{d+2}=iX^0\), which in polar coordinates corresponds to the following combined deformation of $r$ and $\theta$ (see Appendix \ref{Slicing}): 
\begin{equation}
\begin{aligned}
A:\ & \theta = -i\chi,\ \chi: 0 \to \infty,\ \text{and}\ r = i r',\ r': 0 \to \infty,\\
B:\ & \theta = \frac{\pi}{2} + it,\ t: -\infty \to \infty,\\
C:\ & \theta = \pi - i\chi,\ \chi: -\infty \to 0,\ \text{and}\ r = i r',\ r': 0 \to \infty.
\end{aligned}
\label{eq:theta_r_contours}
\end{equation}
---then the contour inevitably crosses the branch cut of complex $r$ plane. As a result, discontinuity contributions must be taken into account, making the phase bookkeeping rather complicated. For this reason, instead of using the rotation \(X^{d+2}=iX^0\), it is more natural and convenient to deform only the \(\theta\) contour, as was done on the L.H.S. of Eq.~\eqref{psi-Phi4}, i.e. from \([0,\pi]\) to the form shown in Eq.~\eqref{eq:theta-contours}.  

In conclusion, also for the R.H.S. of Eq.~\eqref{psi-Phi4}, it is preferable to adopt the prescription of deforming only the \(\theta\) contour, which avoids any assumption about the analytic structure of \(r\).\footnote{Of course, in situations where the external legs are not Mellin transformed---so that the analytic structure of the complex \(r\)-plane is manifest, such as in the usual Euclidean-to-Minkowski continuation of correlators---  the two methods are equivalent, as is explicitly confirmed in Appendix~\ref{Slicing}.}

\section{Connection between Celestial and Cosmological Correlators}\label{sec4}
The celestial dictionary \eqref{celestial_dic} relates the bulk field $\Phi$ in Minkowski spacetime \(\mathrm{M}_{D+2}\) with $\mathcal{O}^\pm_\Delta$ in the celestial  CFT on $\mathrm{S}^{D}$. 
On the other hand, we have related $\Phi$ in \(\mathrm{M}_{D+2}\) with bulk fields $\psi_\lambda$ in de Sitter spacetime \(\mathrm{dS}_{D+1}\). 
Hence, we can relate $\mathcal{O}^\pm_\Delta$ on $\mathrm{S}^{D}$ with $\psi_\lambda$ in \(\mathrm{dS}_{D+1}\). 
More concretely, we consider the late/early limit of the \(\mathrm{dS}\) bulk operator $\psi_\lambda$ that behaves as 
\begin{align}
\label{naive:asympt_behavior_dS_field}
\psi_\lambda \sim e^{\mp\Delta_- t}O_{\Delta_-}^{\pm}+e^{\mp\Delta_+ t}O_{\Delta_+}^{\pm}
\qquad (t\rightarrow \pm\infty),
\end{align}
where `$+$/$-$' denotes late/early limit (or outgoing/incoming), respectively. Recall that $ \Delta_\pm={D}/{2}\pm i\lambda$.
These $O_{\Delta_\pm}^{\pm}$ can be regarded as operators on $\mathrm{S}^{D}$.
In this section, we establish the relation between $\mathcal{O}^\pm_\Delta$ and $O_{\Delta_\pm}^{\pm}$.

Note that the mass of $\psi_\lambda$, given in \eqref{m_lambda}, satisfies $m_\lambda^2 >D^2/4$ and thus \(\psi_\lambda\) is in the principal series of the isometry group of \(\mathrm{dS}_{D+1}\). The GKPW-like dictionary in dS/CFT for the principal series is studied in \cite{Isono:2020qew, Dey:2024zjx}, although their discussions are for the Poincaré patch.
In the GKPW-like dictionary, $O_{\Delta}^{\pm}$ (or their linear combinations) are regarded as the source of a CFT primary in dS/CFT. 
On the other hand, if we suppose a BDHM-like\footnote{The BDHM dictionary \cite{Banks:1998dd} in AdS/CFT relates the asymptotic-boundary limit of the bulk operators to the CFT operators. It is slightly different from the GKPW dictionary \cite{Gubser:1998bc, Witten:1998qj}. 
The GKPW dictionary identifies the asymptotic behavior of non-normalizable modes of bulk fields with the source $J$ of the dual CFT operators. 
We can obtain CFT correlators by differentiating the partition function by $J$. The two dictionaries are equivalent in AdS/CFT, while they are not equivalent in dS/CFT as argued in \cite{Harlow:2011ke}.} dictionary \cite{Witten:2001kn} (see also e.g., \cite{Xiao:2014uea, Goldar:2024crc}), $O_{\Delta}^{\pm}$ (or their linear combinations) are directly related to a CFT primary in dS/CFT. 
Since the dS/CFT dictionary has not been established well, unlike AdS/CFT, 
we do not assume it here, and we just consider $O_{\Delta_\pm}^{\pm}$ as extrapolated operators of dS fields and will argue the implications to dS/CFT later.

We parametrize the coordinates on $\mathrm{M}_{D+2}$ as
\begin{align}
\label{eq:Mink-param}
X(t, r, \bm{p}) &= \left(r\sinh t,\;r\cosh t\;\bm{p}\right),
\end{align}
where $\bm{p}$ is a unit vector on $\mathrm{S}^{D}$.  For convenience, we decompose the vector $X$ into null vectors $p_\pm = (1,\pm\bm{p})$ as
\begin{align}
X = \frac{1}{2}r e^{t}p_+ \;-\;\frac{1}{2}r e^{-t}p_-.
\end{align}

Then, the extrapolated operators in dS are defined as the leading coefficients of the dS bulk field in the late/early limit:
\begin{align}
\label{asympt_behavior_dS_field}
\psi_\lambda(t,\bm{p}) &\sim e^{-\Delta_- t}O_{\Delta_-}^{+}(p_+)+e^{-\Delta_+ t}O_{\Delta_+}^{+}(p_+)
\qquad (t\rightarrow \infty),\\
\psi_\lambda(t,\bm{p}) &\sim e^{\Delta_- t}O_{\Delta_-}^{-}(p_-)+e^{\Delta_+ t}O_{\Delta_+}^{-}(p_-)
\qquad (t\rightarrow -\infty),
\end{align}
where the embedding coordinates of dS is parametrized as $x(t,\bm{p})=(\sinh t, \cosh t \bm{p})$.\footnote{Here, following the discussions in \cite{Witten:2001kn, Strominger:2001pn}, we use the coordinates of the would-be dual space $S^{D}$ such that each point on the future infinity and its antipodal point on past infinity have the same coordinates. For the same reason, we put $-$ into the argument of creation and annihilation operators for the incoming particle in the celestial dictionary \eqref{celestial_dic} as $a^\dagger_{-}(-\omega \bm{p})$.}
Note that both of the terms in the right-hand side are of the same order because $\Delta_\pm={D}/{2}\pm i\lambda$.

We begin by considering the asymptotic limit $t\rightarrow \pm\infty$, where we can use the free–field expansion of $\Phi$. The free–field expansion leads to 
\begin{align}
\psi_\lambda(t,\bm{p})
&= \frac{1}{\sqrt{2\pi}}
  \int \frac{dr}{r}\,r^{\Delta_-}e^{-\epsilon r}\Phi(X) 
\nonumber\\
&= \frac{1}{\sqrt{2\pi}}
  \int \frac{dr}{r}\,r^{\Delta_-}e^{-\epsilon r}
  \int \frac{\omega^{D-1}\,d\omega \,d^D\bm{k}}{2(2\pi)^{D+1}}
  \left(a_{\pm}(\omega \bm{k})e^{i\omega k\cdot X-\epsilon\omega}
       +a_{\pm}^\dagger(\omega \bm{k})e^{-i\omega k\cdot X-\epsilon\omega}\right).
\label{4.4}
\end{align}
Here, $k=(1,\bm{k})$ with $|\bm{k}|=1$  and $d^D\bm{k}$ denotes the integral over $S^D$. 
The UV cutoff factor $e^{-\epsilon\omega}$ is introduced by adding an infinitesimal imaginary part to the time component of $X$ to extract the vacuum.\footnote{For notational simplicity, we use the same symbol $\epsilon$ for the IR regulator of $r$ and the UV cutoff of $\omega$, although they are different positive infinitesimal quantities.}

Next, we separate the integration \eqref{4.4} into two parts, 
\begin{align}
\psi_\lambda(t,\bm{p}):=\psi^{\Delta_-}_\lambda(t,\bm{p})+\psi^{\Delta_+}_\lambda(t,\bm{p}),
\end{align}
where $\psi^{\Delta_-}_\lambda$ represents the contribution from the region where $\bm{k}\neq\pm\bm{p}$ (i.e.\ $\int_{\bm{k}\neq\pm\bm{p}}d^D\bm{k}\cdots$) and $\psi^{\Delta_+}_\lambda$ represents does the one from a neighborhood around $\bm{k}=\pm\bm{p}$ (i.e.\ $\int_{\bm{k}=\pm\bm{p}}d^D\bm{k}\cdots$). 
We will show that both contributions are of the same order in the limit $t\rightarrow \pm\infty$.
Indeed, the former (latter) contribution corresponds to the first (second) term in the asymptotic expansion \eqref{asympt_behavior_dS_field}.

In \cite{Jorstad:2023ajr}, a similar extrapolation is taken. More precisely, they consider a quantity like 
\begin{align}
e^{\Delta_- t}\psi_\lambda(t,\bm{p}) \sim O_{\Delta_-}^{+}(p)+e^{-2i\lambda}O_{\Delta_+}^{+}(p)
\qquad (t\rightarrow +\infty).
\end{align}
Then, the second term is highly oscillating, and only the first term is picked up.
In our prescription, we keep both of the terms in \eqref{4.4}, and the second term corresponds to a collinear contribution.

\subsubsection*{Contribution from \(\bm{k}\neq \pm\bm{p}\)}
We first consider the contribution from the $\bm{k}\neq \pm\bm{p}$ region of the integral \eqref{4.4}.  Noting that
\begin{align}
\omega k\cdot X = \frac{1}{2}r\omega\left(e^tp_+\cdot k - e^{-t}p_-\cdot k\right),
\end{align}
and  $\bm{k}\neq \pm\bm{p}$ implies $p_{\pm}\cdot k\neq 0$, we change the integration variable as
\begin{align}
r = \frac{r'}{\omega e^{\pm t}(-2p_{\pm}\cdot k)}.
\end{align}
We then expand Eq.~\eqref{4.4} for large $|t|$ and keep the leading term proportional to $e^{\mp2t}$.
\begin{align}
\psi^{\Delta_-}_\lambda
&\sim e^{\mp\Delta_- t}\frac{1}{2(2\pi)^{D+3/2}}
  \int\frac{dr'\,d^D\bm{k}\,d\omega}{r'}
    r'^{\Delta_-}e^{-\epsilon r'}\omega^{\Delta_+-1}
    (-2p_{\pm}\cdot k)^{\Delta_+-D}
    \left(a_\pm(\omega \bm{k})e^{\mp\frac{i}{4}r'} + a_\pm^\dagger(\omega \bm{k})e^{\pm\frac{i}{4}r'}\right)
\nonumber\\
&= e^{\mp\Delta_- t}\frac{2^{2\Delta_-}\Gamma(\Delta_-)}{2(2\pi)^{D+3/2}}
  \int d^D\bm{k}\,d\omega\,\omega^{\Delta_+-1}
    (-2p_{\pm}\cdot k)^{\Delta_+-D}
    \left(i^{\mp\Delta_-}a_\pm(\omega \bm{k})+i^{\pm\Delta_-}a_\pm^\dagger(\omega \bm{k})\right).
    \label{eq:4.9}
\end{align}
Thus, in the limit $t\rightarrow +\infty$, we obtain by using celestial holography dictionary \eqref{celestial_dic} with $p=(1, \bm{p})$
\begin{align}
\psi^{\Delta_-}_\lambda(t,\bm{p})
&\sim e^{-\Delta_- t}S_{\Delta_+}
  \frac{2^{-2i\lambda}\Gamma(\Delta_-)}{2^{5/2}\pi^{D+3/2}}
  \left(i^{-\Delta_-}\widetilde{\mathcal{O}^{+}_{\Delta_+}}(p)
       +i^{\Delta_-}\mathcal{O}_{\Delta_-}^{+\prime}(p)\right)
\nonumber\\
&\sim e^{-\Delta_- t}N_{-\lambda}
  \left(i^{-\Delta_-}\widetilde{\mathcal{O}^{+}_{\Delta_+}}(p)
       +i^{\Delta_-}\mathcal{O}_{\Delta_-}^{+\prime}(p)\right),
\end{align}
while for $t\rightarrow-\infty$, we obtain 
\begin{align}
\psi^{\Delta_-}_\lambda(t,\bm{p})
\sim e^{+\Delta_- t}N_{-\lambda}
  \left(i^{-\Delta_-}\mathcal{O}^{-}_{\Delta_-}(p_-)
       +i^{\Delta_-}\widetilde{\mathcal{O}_{\Delta_+}^{-\prime}}(p_-)\right),
\end{align}
with
\begin{align}
N_\lambda = \frac{\Gamma(-i\lambda)}{2^{5/2}\pi^{D/2+3/2}}.
\end{align}
Therefore, comparing the results with \eqref{asympt_behavior_dS_field}, we identify
\begin{subequations}\label{shadow_primary}
\begin{align}
O^{+}_{\Delta_-}(p)
&= N_{-\lambda}\left(i^{-\Delta_-}\widetilde{\mathcal{O}^{+}_{\Delta_+}}(p)
               +i^{\Delta_-}\mathcal{O}_{\Delta_-}^{+\prime}(p)\right),
\\
O^{-}_{\Delta_-}(p)
&= N_{-\lambda}\left(i^{-\Delta_-}\mathcal{O}^{-}_{\Delta_-}(p)
               +i^{\Delta_-}\widetilde{\mathcal{O}_{\Delta_+}^{-\prime}}(p)\right).
\end{align}
\end{subequations}
Using \eqref{eq:Odagger}, it can also be written as
\begin{subequations}\label{wo_shadow_primary}
\begin{align}
O^{+}_{\Delta_-}(p)
&= N_{-\lambda}\left(i^{-\Delta_-}\left(\mathcal{O}_{\Delta_+}^{+\prime}(p)\right)^\dagger
               +i^{\Delta_-}\mathcal{O}_{\Delta_-}^{+\prime}(p)\right),
\\
O^{-}_{\Delta_-}(p)
&= N_{-\lambda}\left(i^{-\Delta_-}\mathcal{O}^{-}_{\Delta_-}(p)
               +i^{\Delta_-}\left(\mathcal{O}_{\Delta_+}^{-}(p)\right)^\dagger\right).
\end{align}
\end{subequations}

\subsubsection*{Contribution from the neighborhood of \(\bm{k}=\pm\bm{p}\)}

Next, we examine the colinear contribution from the neighborhood of \(\bm{k}=\pm\bm{p}\) in the integral \eqref{4.4}.  Assuming that \(\bm{k}\) is close to \(\bm{p}\) (so that \(p_{\mp}\cdot k\neq 0\)), we change the integration variable as
\begin{align}
    r = \frac{r'}{\omega e^{-t}(-p_{\mp}\cdot k/2)}.
\end{align}
Then, the integral \eqref{4.4} becomes
\begin{align}
\psi^{\Delta_+}_\lambda(t,\bm{p})
&\sim e^{\mp\Delta_- t}
  \frac{1}{2(2\pi)^{D+3/2}}
  \int \frac{dr'\,d^D\bm{k}\,d\omega}{r'}\,r'^{\Delta_-}e^{-\epsilon r'}
        \omega^{\Delta_+-1}(-p_{\mp}\cdot k/2)^{-\Delta_-}
\nonumber\\
&\quad\times
  \left[
    a_\pm(\omega \bm{k})\exp\left(\pm\left(-ir'e^{\pm2t}\frac{p_{\pm}\cdot k}{p_{\mp}\cdot k} + ir'\right)\right)
    + a_\pm^\dagger(\omega \bm{k})\exp\left(\pm\left(ir'e^{\pm2t}\frac{p_{\pm}\cdot k}{p_{\mp}\cdot k} - ir'\right)\right)
  \right].
  \label{psi+-int}
\end{align}
We further introduce the variable
\begin{align}
z^2 = -e^{\pm2t}\frac{p_{\pm}\cdot k}{p_{\mp}\cdot k}.
\end{align}
The variable $z$ can be written as
\begin{align}
z = e^{t}\tan\frac{\phi}{2}, 
\end{align}
for $t\to+\infty$, and 
\begin{align}
z = e^{-t}\cot\frac{\phi}{2},
\end{align}
for $t\to-\infty$,  where $\phi$ is the angle between \(\bm{k}\) and \(\bm{p}\).  Under this change of variables, the measure \(d^D\bm{k}\) becomes
\begin{align}
d^D\bm{k}
= \sin^{D-1}\phi \,d\phi\, d\Omega_{D-1}
= 2^De^{\mp Dt}z^{D-1}\left(1 + e^{\mp2t}z^2\right)^{-D}dz\, d\Omega_{D-1}.
\end{align}
Hence, \eqref{psi+-int} can be written as
\begin{align}
\psi^{\Delta_+}_\lambda(t,\bm{p})
&\sim e^{\mp\Delta_+ t}
  \frac{2^D}{2(2\pi)^{D+3/2}}
  \int \frac{dr'\,dz\,d\omega \,d\Omega_{D-1}}{r'}r'^{\Delta_-}e^{-\epsilon r'}
        \omega^{\Delta_+-1}z^{D-1}\left(1 + e^{\mp2t}z^2\right)^{-(D-\Delta_-)}
\nonumber\\
&\quad~~~~\times
  \left[
    a_\pm(\omega \bm{k})\exp\left(\pm(-ir'z^2 + ir')\right)
    + a_\pm^\dagger(\omega \bm{k})\exp\left(\pm(ir'z^2 - ir')\right)
  \right],
\end{align}
where $\bm{k}$ depends on $z$ and other angles.

Expanding in powers of \(e^{\mp2t}z^2\ll1\) and keeping the leading term where the direction $\bm{k}$ localize to $\phi=0$ for $t\to \infty$ and to $\phi=\pi$ for $t\to -\infty$, we obtain
\begin{align}
\psi^{\Delta_+}_\lambda(t,\bm{p})
&\sim e^{\mp\Delta_+ t}
  \frac{2^D\Omega_{D-1}}{2(2\pi)^{D+3/2}}
  \int \frac{dr'\,dz\,d\omega}{r'}r'^{\Delta_-}e^{-\epsilon r'}
        \omega^{\Delta_+-1}z^{D-1}
\nonumber\\
&\quad~~~~~~\times
  \left[
    a_\pm(\pm\omega \bm{p})\exp\left(\pm(-ir'z^2+ir')\right)
    + a_\pm^\dagger(\pm\omega \bm{p})\exp\left(\pm(ir'z^2-ir')\right)
  \right],
  \label{eq:4.17}
\end{align}
where \(\Omega_{D-1}\) is the volume of the \((D-1)\)-dimensional unit sphere.\footnote{We define \(\Omega_0=2\) for $D=1$.}

Finally, performing the integrals over $r^{\prime}$ and $z$ and using the celestial dictionary \eqref{celestial_dic}, we obtain
\begin{align}
\psi^{\Delta_+}_\lambda(t,\bm{p})
\sim e^{-\Delta_+ t}N_\lambda
  \left(i^{-\Delta_+}\mathcal{O}^{+}_{\Delta_+}(p)
       + i^{\Delta_+}\widetilde{\mathcal{O}^{\prime+}_{\Delta_-}}(p)\right)
\end{align}
for \(t\to\infty\), and 
\begin{align}
\psi^{\Delta_+}_\lambda(t,\bm{p})
\sim e^{+\Delta_+ t}N_\lambda
  \left(i^{-\Delta_+}\widetilde{\mathcal{O}^{-}_{\Delta_-}}(p_-)
       + i^{\Delta_+}\mathcal{O}^{\prime-}_{\Delta_+}(p_-)\right)
\end{align}
for $t\to-\infty$.
Comparing the results with \eqref{asympt_behavior_dS_field}, we obtain the identification:
\begin{subequations}\label{shadow_primary2}
\begin{align}
O^{+}_{\Delta_+}(p)
&= N_\lambda
   \left(i^{-\Delta_+}\mathcal{O}^{+}_{\Delta_+}(p)
        + i^{\Delta_+}\widetilde{\mathcal{O}^{\prime+}_{\Delta_-}}(p)\right),
\\
O^{-}_{\Delta_+}(p)
&= N_\lambda
   \left(i^{-\Delta_+}\widetilde{\mathcal{O}^{-}_{\Delta_-}}(p)
        + i^{\Delta_+}\mathcal{O}^{\prime-}_{\Delta_+}(p)\right).
\end{align}
\end{subequations}
Using \eqref{eq:Odagger}, it can also be written as
\begin{subequations}\label{wo_shadow_primary2}
\begin{align}
O^{+}_{\Delta_+}(p)
&= N_\lambda
   \left(i^{-\Delta_+}\mathcal{O}^{+}_{\Delta_+}(p)
        + i^{\Delta_+}\left(\mathcal{O}_{\Delta_-}^{+}(p)\right)^\dagger\right),
\\
O^{-}_{\Delta_+}(p)
&= N_\lambda
   \left(i^{-\Delta_+}\left(\mathcal{O}_{\Delta_-}^{\prime -}(p)\right)^\dagger
        + i^{\Delta_+}\mathcal{O}^{\prime-}_{\Delta_+}(p)\right).
\end{align}
\end{subequations}

\subsubsection*{Hermitian conjugate operators}
We now consider the Hermitian conjugation of the identification rule. 
We should have
\begin{subequations}
\begin{align}
\psi^\dagger_\lambda(t,\bm{p}) &\sim e^{-\Delta_+ t}\left(O_{\Delta_-}^{+}(p_+)\right)^\dagger+e^{-\Delta_- t}\left(O_{\Delta_+}^{+}(p_+)\right)^\dagger
\qquad (t\rightarrow \infty),\\
\psi^\dagger_\lambda(t,\bm{p}) &\sim e^{+\Delta_+ t}\left(O_{\Delta_-}^{-}(p_-)\right)^\dagger+e^{+\Delta_- t}\left(O_{\Delta_+}^{-}(p_-)\right)^\dagger
\qquad (t\rightarrow -\infty).
\end{align}
\end{subequations}
The conjugation is given by\footnote{Note that $i^{\Delta_\pm}=e^{\frac{i \pi D}{4}\mp \frac{\pi \lambda}{2}}$, $i^{-\Delta_\pm}=e^{-\frac{i \pi D}{4}\pm \frac{\pi \lambda}{2}}$, and $(i^{\Delta_\pm})^\ast=i^{-\Delta_\mp}$.}
\begin{subequations}\label{shadow_primary3}
\begin{align}
\{O_{\Delta_+}^{+}(p)\}^\dagger
&= N_{-\lambda}
   \left(i^{-\Delta_-}\mathcal{O}^{+}_{\Delta_-}(p)
        + i^{\Delta_-}\widetilde{\mathcal{O}^{\prime+}_{\Delta_+}}(p)\right),
\\
\{O_{\Delta_+}^{-}(p)\}^\dagger
&= N_{-\lambda}
   \left(i^{-\Delta_-}\widetilde{\mathcal{O}^{-}_{\Delta_+}}(p)
        + i^{\Delta_-}\mathcal{O}^{\prime-}_{\Delta_-}(p)\right),
\\
\{O_{\Delta_-}^{+}(p)\}^\dagger
&= N_\lambda
   \left(i^{-\Delta_+}\widetilde{\mathcal{O}^{+}_{\Delta_-}}(p)
        + i^{\Delta_+}\mathcal{O}^{\prime+}_{\Delta_+}(p)\right),
\\
\{O_{\Delta_-}^{-}(p)\}^\dagger
&= N_\lambda
   \left(i^{-\Delta_+}\mathcal{O}^{-}_{\Delta_+}(p)
        + i^{\Delta_+}\widetilde{\mathcal{O}^{\prime-}_{\Delta_-}}(p)\right).
\end{align}
\end{subequations}

\subsubsection*{Extrapolated operators as CFT operators}
The obtained results [\eqref{shadow_primary}, \eqref{shadow_primary2} and \eqref{shadow_primary3}] indicate that the extrapolated operators $O^{\pm}_{\Delta_{\pm}}$ in dS are linear combinations of the celestial operator and the shadow operator.
For example, we have
\begin{align}
\label{ex:non-local}
    O^{+}_{\Delta_+}(p)
&= N_\lambda
   \left(i^{-\Delta_+}\mathcal{O}^{+}_{\Delta_+}(p)
        + i^{\Delta_+}\widetilde{\mathcal{O}^{\prime+}_{\Delta_-}}(p)\right).
\end{align}
If we trust the celestial dictionary, $\mathcal{O}^{\pm}_{\Delta}$ and $\mathcal{O}^{\prime \pm}_{\Delta}$ are local primary operators in the celestial CFT on $\mathrm{S}^{D}$. Then, their shadow operators are non-local operators by definition \eqref{def:shadow}.
Thus, the celestial dictionary implies that a naive BDHM-type dictionary does not hold in dS/CFT because we cannot regard the extrapolated operators $O^{\pm}_{\Delta_{\pm}}$ directly as local CFT operators.\footnote{\label{foot:def:celhol}If we define $\mathcal{O}'$ without the shadow transformation in \eqref{celestial_dic}, we could regard $O^{\pm}_{\Delta_{\pm}}$ as local operators. However, as commented in footnote~\ref{footO'}, the definition \eqref{celestial_dic} is a natural one.}
In order to extract a local operator, we have to consider a linear combination of $O_{\Delta_{\pm}}$ and the shadow operators $\widetilde{O_{\Delta_{\mp}}}$.
The opposite statement also holds. If the BDHM-like extrapolated dictionary holds in dS/CFT, $O^{\pm}_{\Delta_{\pm}}$ should be regarded as local CFT operators on $\mathrm{S}^{D}$, and then the celestial operators $\mathcal{O}^{\pm}_{\Delta}$ themselves cannot be regarded as local operators. 

Our results are different from \cite{Jorstad:2023ajr} where $\mathcal{O}$ is represented as a linear combination of extrapolated operators in dS and AdS. In our method, $\mathcal{O}$ is represented only by extrapolated operators in dS.
This is due to our construction such that we directly relate fields in Minkowski space to ones in dS as \eqref{psi-Phi4}.

\section{Consistency Check for Correlation Functions}
We have obtained the identification rule [\eqref{shadow_primary}, \eqref{shadow_primary2} and \eqref{shadow_primary3}] between the extrapolated operators $O_\Delta$ and the celestial operators $\mathcal{O}_\Delta$. 
In this section, we compute the late/early cosmological correlators in dS and the celestial amplitudes independently to check the consistency of the identification rule.

\subsection{Two-Point Functions}\label{Twopoint}
In this subsection, we compute the two-point functions in general dimensions.

\subsubsection*{Example: $\langle (O^{+}_{\Delta_{1+}})^\dagger O^-_{\Delta_{2-}}\rangle$}

We first check 
\begin{align}
  \langle \{O^{+}_{\Delta_{1+}}(p_1)\}^\dagger O^-_{\Delta_{2-}}(p_{2})\rangle
  &= i^{-\Delta_{1-}} i^{-\Delta_{2-}}
     N_{-\lambda_1} N_{-\lambda_2}
     \langle \mathcal{O}^+_{\Delta_{1-}}(p_1) \mathcal{O}^-_{\Delta_{2-}}(p_{2})\rangle .
     \label{two-point}
\end{align}
Note that $\mathcal{O}'$ does not appear in the right-hand side because it annihilates the Minkowski vacuum. 

From the dictionary of celestial holography, the following two-point function is computed as :
\begin{align}
\label{OtO}
  \langle \mathcal{O}^+_{\Delta_{1-}}(p_1)\widetilde{\mathcal{O}^-_{\Delta_{2-}}}(p'_{2})\rangle
  &=(2\pi)^{D+1}
    \int\frac{d\omega_1}{\omega_1}\frac{d\omega_2}{\omega_2}
      \omega_1^{\Delta_{1-}}\omega_2^{\Delta_{2+}}
      (2\omega_1)
      \delta^{(D+1)}(\omega_1 \bm{p}_1-\omega_2 \bm{p}'_{2}) \nonumber\\
  &= 2(2\pi)^{D+2}
     \delta(\lambda_1-\lambda_2)
     \delta^{(D)}(p_1-p'_{2}).
\end{align}
Then, applying the shadow transform and including all factors, the right-hand side of \eqref{two-point} is given by
\begin{align}
  &i^{-\Delta_{1-}} i^{-\Delta_{2-}}
     N_{-\lambda_1} N_{-\lambda_2}
     \langle \mathcal{O}^+_{\Delta_{1-}} \mathcal{O}^-_{\Delta_{2-}}\rangle\nonumber\\
  &= 2(2\pi)^{D+2}
     \delta(\lambda_1-\lambda_2)
     i^{-2\Delta_{1-}}
     N_{-\lambda_1}^2
     \frac{1}{S_{\Delta_+}}
     (-2p_1\!\cdot\! p_{2})^{-\Delta_{1-}} \nonumber\\
  &= i^{-2\Delta_{1-}}
     \frac{\Gamma(\Delta_{1-})\Gamma(i\lambda_1)}
          {4\pi^{D/2+1}}\delta(\lambda_1-\lambda_2)
     (-2p_1\!\cdot\! p_{2})^{-\Delta_{1-}}.
\end{align}
This result agrees with the late/early two-point function $\langle O^{+\dagger}_{\Delta_{1+}} O^-_{\Delta_{2-}}\rangle$ obtained in dS which is computed in Appendix~\ref{pro_of_S} [see \eqref{eq:boundary_2pt}].
Note also that the result demonstrates that $\langle (O^{+}_{\Delta_{1+}}(p_1))^\dagger O^-_{\Delta_{2-}}(p_{2-})\rangle$ takes a form of CFT two-point functions with conformal dimension $\Delta_{1-}$.
However, it does not imply that the extrapolated operators $O_\Delta$ behave as CFT primary operators, as remarked around \eqref{ex:non-local}.

\subsubsection*{Example: $\langle (O^{+}_{\Delta_{1+}})^\dagger O^-_{\Delta_{2+}}\rangle$}

Next, we consider the following relation:
\begin{align}
 \langle \{O^{+}_{\Delta_{1+}}(p_1)\}^\dagger O^-_{\Delta_{2+}}(p_{2})\rangle
  &= i^{-\Delta_{1-}} i^{-\Delta_{2+}}
     N_{-\lambda_1} N_{\lambda_2}
     \langle \mathcal{O}^+_{\Delta_{1-}}(p_1)
             \widetilde{\mathcal{O}^-_{\Delta_{2-}}}(p_{2})\rangle .
\end{align}
Using \eqref{OtO}, we find that the right-hand side is
\begin{align}
   &i^{-\Delta_{1-}} i^{-\Delta_{2+}}
     N_{-\lambda_1} N_{\lambda_2}
     \langle \mathcal{O}^+_{\Delta_{1-}}
             \widetilde{\mathcal{O}^-_{\Delta_{2-}}}\rangle\nonumber\\
  &= 2(2\pi)^{D+2}
     \delta(\lambda_1-\lambda_2)
     i^{-\Delta_{1-}-\Delta_{1+}}
     N_{-\lambda_1} N_{\lambda_2}
     \delta^{(D)}(p_1-p_{2})\nonumber \\
  &= 2^{D}
     i^{-D}
     \frac{\Gamma(i\lambda_1)\Gamma(-i\lambda_1)}
          {4\pi}
     \delta(\lambda_1-\lambda_2)
     \delta^{(D)}(p_1-p_{2}).
     \label{eq:5.5}
\end{align}
This result also agrees with the two-point function in dS [see \eqref{eq:boundary_2pt}].
Although one may think that this delta-function behavior (i.e., the contact term) is unusual, it is known \cite{Hogervorst:2021uvp, Sengor:2021zlc, SalehiVaziri:2024joi} that such a contact term arises in the late–late two-point functions. In Appendix~\ref{pro_of_S}, we confirm that the contact term also arises in the late-early correlators.

The result \eqref{eq:5.5} and \eqref{eq:boundary_2pt} also shows that the two-point functions of the extrapolated operators in dS generally do not take the standard power-law form (e.g., $(-2p_1\cdot p_2)^{-\Delta}$). It suggests that it is more natural to interpret it as a linear combination of a primary operator and its shadow counterpart as argued in the previous section.
This perspective has also been suggested in \cite{SalehiVaziri:2024joi}.

\subsection{Four-Point Function ($D=2$, $\phi^{4}$ interaction)}
In this section, we consider the four-point function in $D=2$ with $\phi^{4}$ interaction case at the tree level.
\subsubsection*{Example:  
$\left\langle 
   \left({O}_{\Delta_{1+}}^{+}\right)^\dagger
   {O}_{\Delta_{2+}}^{+}
   {O}_{\Delta_{3+}}^{-}
   {O}_{\Delta_{4+}}^{-} 
\right\rangle$}

We would like to demonstrate the following equation:
\begin{align}
&\left\langle 
   \left({O}_{\Delta_{1+}}^{+}(p_1)\right)^\dagger
   {O}_{\Delta_{2+}}^{+}(p_2)
   {O}_{\Delta_{3+}}^{-}(p_{3})
   {O}_{\Delta_{4+}}^{-} (p_{4})
\right\rangle\nn
  &~~~~~~~~~~~~~~~~~~~~~~~~~~~~=
  i^{-\sum \Delta_i}
  N_{-\lambda_1} N_{\lambda_2} N_{\lambda_3} N_{\lambda_4}
  \left\langle
    \mathcal{O}^{+}_{\Delta_{1-}}(p_1)
    \mathcal{O}^{+}_{\Delta_{2+}}(p_2)
    \widetilde{\mathcal{O}^{-}_{\Delta_{3-}}}(p_{3})
    \widetilde{\mathcal{O}^{-}_{\Delta_{4-}}}(p_{4})
  \right\rangle .
  \label{eq:4pt-tbp}
\end{align}

In the following, we use $\Delta_i$ to denote the scaling dimension associated with the $i$-th operator under the rescaling $p_i \rightarrow a p_i$. Specifically, since $(O_{\Delta_{\pm}}(p))^\dagger$ transforms as  
\begin{align}
   (O_{\Delta_{\pm}}(p))^\dagger \;\;\longrightarrow\;\; a^{-\Delta_{\mp}} \,(O_{\Delta_{\pm}}(p))^\dagger ,
\end{align}
we define $\Delta_i \equiv \Delta_{i\mp}$ for the conjugate operator $\left(O_{\Delta_{i\pm}}(p_i)\right)^\dagger$. 
Similarly, for the shadow operator $\widetilde{{O}_{\Delta_{i\pm}}}(p_i)$, we also define $\Delta_i \equiv \Delta_{i\mp}$.

\paragraph{Left-hand side}

The left-hand side (LHS) is evaluated from a QFT with the interaction \eqref{L_int} on $S^{D+1}$ through the analytic continuation explained in section~\ref{sec3}.  
The external vertices on dS are obtained from the sphere by the continuation \eqref{theta_i-t_i}. 
Furthermore, by deforming the path-integral contour, the interaction vertex moves into the region $\mathrm{EAdS}^{+}+\mathrm{dS}+\mathrm{EAdS}^{-}$ as in \eqref{eq:theta-contours}, and we can apply formula \eqref{formula_deform} when performing the contour deformation.  
Carrying out this continuation, we obtain\footnote{Prefactor $i^{-D-2}$ of $\int_{\EAdS}$ is  $1$ since now we consider $D=2$.}
\begin{align}
 &\left\langle
   (O_{\Delta_{1+}}^{+})^\dagger
   O_{\Delta_{2+}}^{+}
   O_{\Delta_{3+}}^{-}
   O_{\Delta_{4+}}^{-}
 \right\rangle\nonumber\\
 &=
 -i \alpha_4 \, 2\pi \,
 \delta(-\lambda_1+\lambda_2+\lambda_3+\lambda_4) 
 \left(
   \int_{\text{EAdS}^{+}} [d^{3}y]
   + \int_{\text{EAdS}^{-}} [d^{3}y]
   + \int_{\text{dS}} [d^{3}y]
 \right)\nonumber
 \\
 &~~~~~~~~~~~~~~~~~~~~~\times
 \prod_{i=1}^{4}
   \frac{1}{\sqrt{2\pi}}
   \frac{2^{\Delta_i}
         \Gamma(h-\Delta_i)
         \Gamma(\Delta_i)}
        {4\pi^{2}}
   \left[{-}2\eta_i( p_i\!\cdot\!x)+i\epsilon\right]^{-\Delta_i},
   \label{eq:5.7}
\end{align}
where $x:=-iy$ if $y\in \EAdS$ and $x:=y$ if $y\in \dS$, $\alpha_4$ is the coupling constant defined in \eqref{L_int}, and $\eta_i=\pm 1$ specifies whether the operator is inserted at the future boundary $\eta=+1$ or the past boundary $\eta=-1$.

 As already mentioned in Sec.~\ref{sec3}, we emphasize once again that the contribution from EAdS arises purely for computational convenience to prepare the Bunch-Davies vacuum, 
and does not imply that we are considering a disconnected spacetime obtained by gluing EAdS and dS together. 
In our construction, EAdS appears as a result of deforming the integration contour for internal vertices.
The appearance of an EAdS integral is merely a technical artifact of working with the Bunch-Davies vacuum in dS. 
Therefore, one should keep in mind that our discussion consistently concerns dS spacetime itself.

\paragraph{Right-hand side}

The corresponding Lorentzian four-point celestial amplitude is obtained by applying Mellin transforms to the tree-level amplitude in four-dimensional Minkowski spacetime, which is reached via analytic continuation of the Euclidean correlator. In this process, as in the relevant equations, we can apply formula \eqref{formula_deform} to the angular integration over the internal vertex.

\begin{align}
 \left\langle
   \mathcal{O}^{+}_{\Delta_{1-}}
   \mathcal{O}^{+}_{\Delta_{2+}}
   \widetilde{\mathcal{O}^{-}_{\Delta_{3-}}}
   \widetilde{\mathcal{O}^{-}_{\Delta_{4-}}}
 \right\rangle
  &=
 -i\alpha_4 \int dr \, r^{3}
 \left(
   \int_{\text{EAdS}^{+}} [d^{3}y] +
   \int_{\text{EAdS}^{-}} [d^{3}y] +
   \int_{\text{dS}} [d^{3}y]
 \right)\nonumber\\
 &\qquad\qquad\times
 \int \!\left(\prod_{i=1}^{4} d\omega_i \,\omega_i^{\Delta_i-1}
             e^{-\epsilon \omega_i}
             e^{-i\eta_i \omega_i r \, p_i\!\cdot\! x}\right)\nonumber\\
  &=
 -i\alpha_4 \, 2\pi \,
 \delta(-\lambda_1+\lambda_2+\lambda_3+\lambda_4)\,
 \left(
   \int_{\text{EAdS}^{+}} [d^{3}y] +
   \int_{\text{EAdS}^{-}} [d^{3}y] +
   \int_{\text{dS}} [d^{3}y]
 \right)\nonumber\\
 &\qquad\qquad\times
 \prod_{i=1}^{4} (2i)^{\Delta_i}\Gamma(\Delta_i)
 \left[{-}2\eta_i (p_i\!\cdot\! x)+i\epsilon\right]^{-\Delta_i}.
 \label{eq:4pt-celes}
\end{align}
Here, from the first to the second equality, we performed the change of variables
$\omega_i \to \omega_i/r$.

\paragraph{Matching of both sides}

Multiplying \eqref{eq:4pt-celes} by the overall factors, we obtain
\begin{align}
  &i^{-\sum \Delta_i}
  N_{-\lambda_1} N_{\lambda_2} N_{\lambda_3} N_{\lambda_4}
  \left\langle
    \mathcal{O}^{+}_{\Delta_{1-}}
    \mathcal{O}^{+}_{\Delta_{2+}}
    \widetilde{\mathcal{O}^{-}_{\Delta_{3-}}}
    \widetilde{\mathcal{O}^{-}_{\Delta_{4-}}}
  \right\rangle\nonumber\\
  &=-i\alpha_4 2\pi 
   \delta(-\lambda_1+\lambda_2+\lambda_3+\lambda_4)
   \left(
     \int_{\text{EAdS}^{+}}\! [d^{3}y] +
     \int_{\text{EAdS}^{-}}\! [d^{3}y] +
     \int_{\dS}\! [d^{3}y]
   \right)
   \nn
   &\qquad \qquad \times
   \prod_{i=1}^{4}
     2^{\Delta_i} N_{\lambda_i} \Gamma(\Delta_i)
     \left[ {-}2\eta_i (p_i\!\cdot\!x)+i\epsilon\right]^{-\Delta_i}.
\end{align}
Since we have 
\begin{align}
  2^{\Delta_i}N_{\lambda_i}\Gamma(\Delta_i)
  = \frac{1}{\sqrt{2\pi}} 
    \frac{2^{\Delta_i}\Gamma(1-\Delta_i)\Gamma(\Delta_i)}
         {4\pi^{2}},
\end{align}
we confirm that \eqref{eq:4pt-tbp} holds.

\subsubsection*{Example:  
$\left\langle 
   \left({O}_{\Delta_{1+}}^{+}\right)^\dagger
   {O}_{\Delta_{2+}}^{+}
   {O}_{\Delta_{3+}}^{-}
   {O}_{\Delta_{4+}}^{-} 
\right\rangle$}
In the above example \eqref{eq:4pt-tbp}, no contribution from the shadow conformal primary wave function appeared. 
Here we consider a case where such a contribution does arise. 
Specifically, let us consider modifying the second operator in the previous example as  
\begin{align}
&\left\langle 
   \left({O}_{\Delta_{1+}}^{+}(p_1)\right)^\dagger
   {O}_{\Delta_{2-}}^{+}(p_2)
   {O}_{\Delta_{3+}}^{-}(p_{3})
   {O}_{\Delta_{4+}}^{-} (p_{4})
\right\rangle\nn
  &~~~~~~~~~~~~~~~~~~~~~~~~~~~~=
  i^{-\sum \Delta_i}
  N_{-\lambda_1} N_{-\lambda_2} N_{\lambda_3} N_{\lambda_4}
  \left\langle
    \mathcal{O}^{+}_{\Delta_{1-}}(p_1)
    \widetilde{\mathcal{O}^{+}_{\Delta_{2+}}}(p_2)
    \widetilde{\mathcal{O}^{-}_{\Delta_{3-}}}(p_{3})
    \widetilde{\mathcal{O}^{-}_{\Delta_{4-}}}(p_{4})
  \right\rangle .
  \label{eq:4pt-shadow}
\end{align}
First, the expression on the L.H.S. of \eqref{eq:4pt-shadow} is
\begin{align}
&\left\langle 
   \left({O}_{\Delta_{1+}}^{+}(p_1)\right)^\dagger
   {O}_{\Delta_{2-}}^{+}(p_2)
   {O}_{\Delta_{3+}}^{-}(p_{3})
   {O}_{\Delta_{4+}}^{-} (p_{4})
\right\rangle \nn
&=-i\alpha_4  2\pi
 \delta(-\lambda_1+\lambda_2+\lambda_3+\lambda_4)\; 
 \left(
   \int_{\text{EAdS}^{+}}\! [d^{3}y]
   + \int_{\text{EAdS}^{-}}\! [d^{3}y]
   + \int_{\dS}\! [d^{3}y]
 \right)\nonumber
 \\
 &~~~~~~~~~\quad\times
 \prod_{i=1}^{4}
   \frac{1}{\sqrt{2\pi}}
   \frac{2^{\Delta_i}
         \Gamma(h-\Delta_i)
         \Gamma(\Delta_i)}
        {4\pi^{2}}
   \left[{-}2\eta_i(p_i\!\cdot\!x)+i\epsilon\right]^{-\Delta_i}.
\end{align}
Second, we consider the R.H.S. of \eqref{eq:4pt-shadow}. 
 By applying the shadow transform with respect to $p_2$, using \eqref{shadow-formula1} on both sides of \eqref{eq:4pt-celes}, one obtains the R.H.S. of \eqref{eq:4pt-shadow}.\footnote{Here we note that, in our notation, in \eqref{eq:4pt-tbp} we defined $\Delta_2 \equiv \Delta_{2+}$, whereas in \eqref{eq:4pt-shadow} we defined $\Delta_2 \equiv \Delta_{2-}$.}
\begin{align}
&\left\langle
    \mathcal{O}^{+}_{\Delta_{1-}}(p_1)
    \widetilde{\mathcal{O}^{+}_{\Delta_{2+}}}(p_2)
    \widetilde{\mathcal{O}^{-}_{\Delta_{3-}}}(p_{3})
    \widetilde{\mathcal{O}^{-}_{\Delta_{4-}}}(p_{4})
  \right\rangle \nn
  &=-i\alpha_4 2\pi
 \delta(-\lambda_1+\lambda_2+\lambda_3+\lambda_4)\; 
 \left(
   \int_{\text{EAdS}^{+}}\! [d^{3}y] +
   \int_{\text{EAdS}^{-}}\! [d^{3}y] +
   \int_{\dS}\! [d^{3}y]
 \right)\nonumber\\ 
 &\times
 (2i)^{\Delta_{2+}}\Gamma(\Delta_{2+})\frac{1}{S_{\Delta_{2+}}}\frac{\pi^h\Gamma(i\lambda_2)}{\Gamma(\Delta_{2+})}
 (-1+i\epsilon)^{-i\lambda_2}\left[{-}2(p_2\!\cdot\!x)+i\epsilon\right]^{-\Delta_2}\prod_{i=1,3,4} (2i)^{\Delta_i}\Gamma(\Delta_i)
 \left[{-}2\eta_i(p_i\!\cdot\!x)+i\epsilon\right]^{-\Delta_i}\nn
 &=-i\alpha_4 2\pi
 \delta(-\lambda_1+\lambda_2+\lambda_3+\lambda_4)\; 
 \left(
   \int_{\text{EAdS}^{+}}\! [d^{3}y] +
   \int_{\text{EAdS}^{-}}\! [d^{3}y] +
   \int_{\dS}\! [d^{3}y]
 \right)\nonumber\\ 
 &\times
 \prod_{i=1}^{4} (2i)^{\Delta_i}\Gamma(\Delta_i)
 \left[{-}2\eta_i(p_i\!\cdot\!x)+i\epsilon\right]^{-\Delta_i}.
\end{align}
Therefore, we confirm that our correspondence also holds for this case. 

At this point, the reader might suspect that one can replace the insertion of $\mathcal{O}^+_{\Delta_-}$ 
with that of $\widetilde{\mathcal{O}^+_{\Delta_+}}$. 
However, it is important to note that this replacement changes the argument of the delta function from $-\lambda$ to $\lambda$. 
This provides another way to see why the operator corresponding to $(O^+_{\Delta_+})^\dagger$ must be $\mathcal{O}^+_{\Delta_-}$, 
while the operator corresponding to $O^+_{\Delta_-}$ must be $\widetilde{\mathcal{O}^+_{\Delta_+}}$.

\section{Conclusion}
\label{sec:sumdis}
\subsection{Summary and Discussions}

In this paper, we have started by considering a massless scalar field on flat space 
\(\mathbb{R}^{D+2}\). Through a Weyl transformation and a Fourier transformation along a radial direction, we have mapped this theory to that of massive
scalar fields on the sphere \(\mathrm{S}^{D+1}\). 
Under an analytic continuation, the theory on \(\mathrm{S}^{D+1}\) is mapped to the theory in \(\mathrm{dS}_{D+1}\) with the Bunch--Davies 
vacuum. The original \(\mathbb{R}^{D+2}\) theory is also mapped to 
Minkowski space \(\mathrm{M}_{D+2}\) via the analytic continuation. 
We therefore have observed a relation between correlation functions on Minkowski space and de Sitter space.

This identification provides a chance to relate the celestial holography with the dS/CFT correspondence.
To explore this possibility, we have examined how operators $\mathcal{O}_\Delta$ on the celestial sphere are related to extrapolated operators $O_\Delta$ in dS by assuming a dictionary of the celestial holography.
The obtained relation among these operators is
\eqref{shadow_primary}, \eqref{shadow_primary2} and \eqref{shadow_primary3}.
We have also confirmed the relation by computing the celestial amplitudes and the cosmological correlators independently.
The identification rules enable us to represent the scattering amplitudes in the Minkowski space $\mathrm{M}_{D+2}$ and the cosmological correlators in $\mathrm{dS}_{D+1}$ as correlators on $\mathrm{S}^{D}$. 

Celestial holography conjectures that the hypothetical theory on $\mathrm{S}^{D}$ is a conformal field theory with the primary operators $\mathcal{O}_\Delta$. 
Our work indicates that this CFT can also describe the cosmological correlators for $\mathrm{dS}_{D+1}$.
This provides implications for the dictionary of the dS/CFT correspondence, which remains far less settled than AdS/CFT.
In particular, our work shows that the celestial holography implies that a naive BDHM-type dictionary in dS/CFT does not hold,\footnote{As commented in footnote~\ref{foot:def:celhol}, this conclusion depends on the precise definition of the celestial dictionary \eqref{celestial_dic}.} because $O_\Delta$ is given by a linear combination of $\mathcal{O}_\Delta$ and the non-local shadow operator $\widetilde{\mathcal{O}_{D-\Delta}}$.
Although further work is required to relate our construction to the conventional dictionary of dS/CFT---such as a GKPW-type dictionary commented in the next subsection---, we expect that our identification between theories on flat space and dS offers a fresh perspective on the origin of the dictionary for celestial holography and dS/CFT.

There are many open problems in celestial holography. 
Our framework may shed new light on them by importing methods and ideas developed in dS/CFT (for instance, central extensions) and as well as techniques in the cosmological bootstrap (see, e.g., \cite{Benincasa:2022gtd}).
In addition, a relation between flat holography and AdS/CFT has also been developed by taking a flat-space limit of AdS/CFT \cite{Polchinski:1999ry, Susskind:1998vk, Giddings:1999jq}. 
Furthermore, recent progress \cite{Sleight:2020obc, DiPietro:2021sjt, Sleight:2021plv, Loparco:2023rug} has revealed several bridges between dS and AdS. 
Motivated by (A)dS/CFT, an extrapolate dictionary for celestial holography is also proposed \cite{Iacobacci:2022yjo, Sleight:2023ojm, Jorstad:2023ajr, Iacobacci:2024nhw}.
Combining our results with these works, we may treat (A)dS/CFT and flat holography on a comparable footing, and can find a common platform for the holographic description of quantum gravity.

\subsection{Future directions}

\begin{enumerate}
\item{\textbf{dS/CFT dictionary.}}
Unlike AdS/CFT, the GKPW-type dictionary \cite{Strominger:2001pn, Maldacena:2002vr} is not equivalent to the BDHM-type extrapolate dictionary in dS/CFT, as argued in \cite{Harlow:2011ke}. 
In the GKPW-type dictionary of dS/CFT, the extrapolated operators $O_\Delta$ are regarded as the source of primary operators for the dual CFT (see \cite{Isono:2020qew}),
and the source is regarded as an argument of the bulk wave function.
We have examined the relation to the extrapolated operators $O_\Delta$ to the celestial operators $\mathcal{O}_\Delta$ and have not specifed the dictionary in dS/CFT. 
While it seems that the BDHM-type naturally fits with the celestial dictionary (though $O_\Delta$ itself are not CFT local operators as mentioned around \eqref{ex:non-local}), 
it would be interesting to explore the implication of the GKPW-type interpretation to celestial holography.

\item{\textbf{Other conformal dimension.}}  
Exploring operators beyond the principal series $\Delta=D/2+i\lambda$ is important in both dS/CFT and celestial holography. 
In our construction of theories on $\mathrm{dS}_{D+1}$ from Minkowski spacetime $\mathrm{M}_{D+2}$, scalar fields belong to the principal series with masses satisfying $m^2 > D^2/4$ as in \eqref{m_lambda}.
Light fields with $m^2 < D^2/4$, corresponding to the complementary series, do not appear.
Since light fields in dS play a crucial role in phenomenology and cosmological applications, this issue deserves particular attention.  
In the context of celestial holography, conformally soft operators correspond to integer values of $\Delta$.
It is known that the principal series forms a complete basis \cite{Donnay:2020guq}.
In principle, other operators -- for example, those with integer $\Delta$ -- can be expressed as linear combinations of operators in the principal series (see also \cite{Freidel:2022skz, Cotler:2023qwh}).
It would be interesting to carry out this analysis and determine what the soft operators in celestial holography correspond to in de Sitter space.


\item \textbf{Extension to the massive case.}
  In this work, we have focused on the massless scalar field in Minkowski spacetime. 
  If we introduce a mass term as
  \begin{equation}
      S[\Phi]=\int_{\mathbb{R}^{D+2}} d^{D+2} X\left[\frac{1}{2} \partial_I \Phi \partial^I \Phi+\frac{1}{2} m^2 \Phi^2+\frac{D}{8(D+1)} R[G] \Phi^2\right],
  \end{equation}
  the resulting action on $S^{D+1}$ is
  \begin{align}      
    \begin{aligned}
    S[\psi]= \frac{1}{2}&\int_{S^{D+1}} \qty[d^{d+1} x] \iint_{-\infty}^{\infty} d \lambda d \lambda^{\prime} ~\Bigl[\delta(\lambda+\lambda^{\prime})g_S^{ij}\nabla_i\psi_\lambda(x) \nabla_j\psi_{\lambda^{\prime}}(x)\Bigr. \\
    & ~~~~~~~~~~~~\bigl.+\left\{\left(\lambda^2+\frac{D^2}{4}\right) \delta\left(\lambda+\lambda^{\prime}\right)
    +m^2 \delta^G\left(-\lambda-\lambda^{\prime}+2 i\right)\right\}\psi_\lambda(x)\psi_{\lambda^{\prime}}(x)\Bigr].
    \end{aligned}
  \end{align}
Thus, the action is not diagonal with respect to $\lambda$, and the analysis becomes more complicated than the massless case.
  
  In addition, we also note that the celestial dictionary for the massive field is no longer simply expressed by the Mellin transform as \eqref{celestial_dic}, and we have to use a bulk-boundary propagator in Euclidean AdS (see \cite{Pasterski:2016qvg, Pasterski:2017kqt, Furugori:2023hgv}).
  Note that for our analysis on massless fields, the Mellin transformation appeared after the saddle point approximation in \eqref{eq:4.9} and \eqref{eq:4.17}. 
  We expect that the bulk-boundary propagator also appears in the saddle point approximation for massive fields. Indeed, to see the asymptotic behavior of the massive fields, it is natural to use a hyperbolic parametization of the on-shell momentum (see, e.g.,  \cite{Campiglia:2015qka, Hirai:2018ijc}).

\item \textbf{Generalization to gauge theories and gravity: currents, stress tensors, and soft theorems.}
  Beyond scalar fields, it is important to extend a similar analysis to gauge and gravitational theories.  
  In the celestial holography, the gauge currents and stress tensors are closely 
  tied to soft theorems \cite{Kapec:2016jld}, and a feature looks different from (A)dS/CFT \cite{Balasubramanian:2001nb, Balasubramanian:1999re, Henningson:1998gx}. 
  It would be interesting to see whether the soft currents in celestial holography can be derived from a (A)dS/CFT-type construction, and we are now carrying out a detailed analysis.

\item \textbf{\(D=0\) case.}
  If we consider two-dimensional Minkowski spacetime (i.e., setting \(D=0\) in our analysis), the corresponding de Sitter space is one-dimensional, and 
  it seems that our methods cannot be applied straightforwardly. Nonetheless, two-dimensional gravitational 
  models such as Liouville theory \cite{Polyakov:1981rd} are solvable and of considerable interest, and we remark that celestial amplitudes for two-dimensional Minkowski are investigated in \cite{Duary:2022onm}.
  It remains an interesting open question to adapt our framework to two-dimensional settings.

\end{enumerate}

\section*{Acknowledgements}
We are grateful to Shinji Mukohyama and Tadashi Takayanagi, Takahiro Tanaka and Yu-ki Suzuki for useful discussions.
HF is supported by JSPS Grant-in-Aid for Scientific Research KAKENHI Grant No. JP22H01217.
NO is supported by JSPS KAKENHI Grant Number JP24KJ1372. 
TW is supported by JSPS KAKENHI Grant Number JP25KJ1621. 
The work of NO is partially supported by Grant-inAid for Transformative Research Areas (A) “Extreme Universe” No. 21H05187.
SS acknowledges support from JSPS KAKENHI Grant Numbers JP21K13927 and JP22H05115, and JST BOOST Program Japan Grant Number JPMJBY24E0.

\appendix
\section{Two-point Functions on $S^2$ from $\mathbb{R}^{3}$}\label{2dsphere}
From \eqref{eq:psi-Phi} for $D=1$, 
the two-point functions on $\mathbb{R}^{3}$ and $S^2$ should be related as
\begin{align}
    \langle \psi^\dagger_{\lambda_1}(x_1) \psi_{\lambda_2}(x_2) \rangle 
    &= \frac{1}{2\pi}\int d\xi_1   d\xi_2   e^{(1/2 + i\lambda_1)\xi_1}   e^{(1/2 - i\lambda_2)\xi_2}  \langle \Phi(X_1) \Phi(X_2) \rangle \nonumber\\
    &= \frac{1}{8\pi^2}\int d\xi_1   d\xi_2   e^{(1/2 + i\lambda_1)\xi_1}   e^{(1/2 - i\lambda_2)\xi_2} \frac{1}{\sqrt{ e^{2\xi_1} + e^{2\xi_2} - 2 e^{\xi_1+\xi_2} \cos\gamma }},
\end{align}
where \(\gamma\) is the angle between the two points $X_1, X_2$. 

We will confirm that the right-hand side indeed reproduces the two-point function for the Lagrangian \eqref{LofS} on $S^2$.
Performing the change of variables \(u = (\xi_1 - \xi_2)/2\) and \(v = (\xi_1 + \xi_2)/2\), it can be written as
\begin{align}
\langle \psi_{\lambda_1}^\dagger(x_1) \psi_{\lambda_2}(x_2) \rangle 
&= \frac{1}{4\pi^2} \int du  dv   e^{iu(\lambda_1 + \lambda_2)}  e^{ iv(\lambda_1 - \lambda_2)} \frac{1}{\sqrt{e^{2u} + e^{-2u} - 2\cos\gamma}}\nonumber\\
&=\frac{1}{2\pi} \delta(\lambda_1 - \lambda_2)\int du   e^{2iu\lambda_1}\frac{1}{\sqrt{e^{2u} + e^{-2u} - 2 \cos\gamma}}.
\end{align}

We can rewrite this correlation function in the following form by partially expanding it.\footnote{Recall the standard three-dimensional multipole (Legendre) expansion:
\begin{align}
\frac{1}{\sqrt{ r^{2} + r'^{2} - 2rr'\cos\theta}}
\;=\;
\sum_{l=0}^{\infty}
  \frac{r_{<}^{ l}}{r_{>}^{ l+1}}
  P_{l}\left(\cos\theta\right),
\end{align}
where \(r_{<} = \min(r,r')\) and \(r_{>} = \max(r,r')\).}
\begin{align}
\langle \psi_{\lambda_1}^\dagger(x_1) \psi_{\lambda_2}(x_2) \rangle 
&= \frac{1}{2\pi}\delta(\lambda_1 - \lambda_2)\int du\,   e^{2iu\lambda_1} \sum_{l=0}^{\infty} e^{-(2l+1)|u|} P_{l}(\cos\gamma)\nonumber\\
&= \frac{1}{\pi}\delta(\lambda_1 - \lambda_2)\sum_{l=0}^{\infty}
  \frac{(2l+1)P_{l}\left(\cos\gamma\right)}{(2l+1)^{2} + 4\lambda_1^{2}},\nonumber\\
&= \delta(\lambda_1 - \lambda_2)\frac{1}{4\pi}\sum_{l=0}^{\infty}
  \frac{(2l+1)P_{l}\left(\cos\gamma\right)}{l(l+1)+m_{\lambda_1}^2}\nonumber\\
&=\delta(\lambda_1 - \lambda_2)G(x_1,x_2),
\end{align}
where $G(x_1,x_2)$ is the two-point function for the massive scalar field on $S^2$ as reviwed below. The last expression coincides with the two-point function of Lagrangian \eqref{LofS}.

We finally confirm that Green's function on $S^2$ takes
\begin{align}
\label{eq:A4}
G(\gamma) \;=\; \frac{1}{4\pi} \sum_{l=0}^{\infty}  \frac{2l+1}{l(l+1) + m^2} P_l(\cos\gamma).
\end{align}
Consider a massive scalar field with mass $m$ on the two-dimensional sphere \(\mathrm{S}^2\). The Green's function \(G(x,x')\) satisfies
\begin{align}
-(\Delta - m^2) G(x_1,x_2) \;=\; \frac{1}{\sqrt{g_S}} \delta^{(2)}(x_1 - x_2),
\end{align}
where \(\Delta\) is the Laplacian on \(\mathrm{S}^2\) and \(g_S\) is the determinant of its metric.

Since the spherical harmonics \(Y_{lm}(x)\) satisfy
\begin{align}
\Delta Y_{lm} \;=\; - l(l+1) Y_{lm},
\end{align}
one can expand the Green's function as
\begin{align}
G(x_1,x_2) \;=\; \sum_{l=0}^{\infty}\sum_{m^{\prime}=-l}^{l} \frac{Y_{lm^{\prime}}(x) Y_{lm^{\prime}}^*(x')}{ l(l+1)+m^2}.
\end{align}
Using the addition theorem for spherical harmonics,
\begin{align}
\sum_{m=-l}^{l}   Y_{lm}(x_1) Y_{lm}^*(x_2) \;=\; \frac{2l+1}{4\pi} P_{l}(\cos\gamma),
\end{align}
where $\cos\gamma = x_1\cdot x_2$, we obtain \eqref{eq:A4}.

Note that $G$ can also be written as\footnote{It can be confirmed as follows:
\begin{align}
G(x_1,x_2)
\;=\;
\frac{1}{8\pi}
\sum_{l=-\infty}^{\infty}
\frac{2l+1}{l(l+1)+m^2} 
(-1)^l 
P_l\left(-\cos\gamma\right)
\;=\;
\frac{1}{16\pi i} 
\oint_C
dz 
\frac{f(z)}{\sin(\pi z)},
\end{align}
where \(C\) is a contour enclosing all integer poles of $1/\sin(\pi z)$ and $f(z):=
\frac{2z+1}{z(z+1) + m^2} 
P_z\left(-\cos\gamma\right)$.
Picking up the poles $z=-\Delta_\pm$ of $f(z)$, we obtain \eqref{G(gam)=F}.
} 
\begin{align}
\label{G(gam)=F}
    G(x_1,x_2)
\;=\;
\frac{\Gamma\left(\Delta_+\right) \Gamma\left(\Delta_-\right)}{4\pi} 
{}_2F_1\left(\Delta_+,\Delta_-; 1; \frac{1+\cos\gamma}{2}\right)
\end{align}
with $\Delta_\pm = 1/2 \pm i \sqrt{m^2-1/4}$.

\section{Evaluating Integral on $S^{D+1}$ by Contour Deformation of the Polar Angle}
\label{app:contour-split}

Throughout this appendix we work in an \((D{+}2)\)-dimensional
Euclidean space with coordinates
\((X^1,\ldots ,X^{D+1},X^{D+2})=rx\),
and this $x$ denotes the embedding coordinates of $S^{D+1}$.

We would like to consider the following integral
\begin{align}
\label{I:f}
    I=\int_{S^{D+1}} [d^{D+1}x]  f(x),
\end{align}
where $f$ is an arbitary function with an appropriate decaying propoerty in the infinity, and deform the contour of the polar angle of $S^{D+1}$ as in Fig.~\ref{fig:theta-contours} and Eq.\eqref{eq:theta-contours}.

We introduce the polar angle $\theta$ as 
\begin{equation}
  x^{D+2}=-\cos\theta,
  \qquad
  \bm{x}=(x^1,\ldots ,x^{D+1}) = \sin\theta \,\hat n,
  \qquad
  \hat n\in S^{D}.
  \label{eq:spherical-coords}
\end{equation}
The Euclidean volume element is
$[d^{ D+1}x]=(\sin\theta)^{D} d\theta d^D\hat{n}$.

In section~\ref{sec3} we perform the analytic continuation $\theta = \pi/2 + it$. Then, the above coordinates become
\begin{align}
      x^{D+2} \to  i \sinh t =: i x^0,
  \qquad
  \bm{x} \to  \cosh t\, \hat n.
\end{align}
This $(x^0, \bm{x})$ with $-(x^0)^2+ |\bm{x}|^2=1$ represents the embedding of the global coordinates in $\mathrm{dS}_{D+1}$.

For the continuation $\theta= -i\chi$, $x$ becomes as
\begin{align}
      x^{D+2} \to  -\cosh \chi =:-iy^{D+2}= y^{0},
  \qquad
  \bm{x} \to  -i\sinh \chi\, \hat n =: -i\bm{y} .
\end{align}
This $(\bm{y}, y^{0})$ with $-(y^{0})^2+ |\bm{y}|^2=-1$ represents the embedding of the global coordinates in $\mathrm{EAdS}^{-}_{D+1}$ (i.e., $(D+1)$-dimensional unit hyperbolic space $H^-$).

For the continuation $\theta= \pi-i\chi$, $x$ becomes as
\begin{align}
      x^{D+2} \to  \cosh \chi =: -iy^{D+2}= y^{0},
  \qquad
  \bm{x} \to i\sinh \chi\, \hat n =: -i\bm{y} .
\end{align}
This $(\bm{y}, y^{0})$ with $-(y^{0})^2+ |\bm{y}|^2=-1$ represents the embedding of the global coordinates in $\mathrm{EAdS}^{+}_{D+1}$ (i.e.,  hyperbolic space $H^+$).

For a sufficiently rapidly–decaying test function
$f\!\left(x^1,\dots ,x^{D+2}\right)$ we start from
\begin{align}
  I &\;=\;
  \int_{S^{D+1}}\![d ^{D+1}x]\,
  f\!\left(\sin\theta \,\hat n,\;-\cos\theta\right).\nonumber\\
  &=
  \int_{S^{D}}\![d ^D\hat{n}]
  \int_{0}^{\pi}\!d\theta 
  \sin^D\theta \,
  f\!\left(\sin\theta\, \hat n,\;-\cos\theta\right).
  \label{eq:IE}
\end{align}
The real segment \(\theta\in[0,\pi]\) can be deformed as shown in Eq.\eqref{eq:theta-contours} and Fig.\ref{fig:theta-contours} because the remaining horizontal pieces lie at
\(\operatorname{Im}\theta ={\pm}\infty\) do not contribute, assuming
\(f\) decays rapidly there and we also assume that $f$ has an appropriate analytical property on $\theta$ plane so that we can deform the contour safely. Hence, we can write \(I=I_{A}+I_{B}+I_{C}\) with
\begin{subequations} \label{eq:IACE}
\begin{align}
  I_{A} &=
  i^{-D-1}
  \!\int_{S^{D}}\![d^D\hat{n}]
  \!\int_{0}^{\infty}\!d\chi\;
  \sinh^D\chi \,
  f\!\left(-i \sinh\chi\, \hat n,\;-\cosh\chi\right),\\
  I_{B} &=
   i
  \!\int_{S^{D}}\![d^D\hat{n}]
  \!\int_{-\infty}^{\infty}\!dt\;
  \cosh^D t \,
  f\!\left(\cosh t\, \hat n,\;i \sinh t\right),
  \\
  I_{C} &=
  i^{-D-1}
  \!\int_{S^{D}}\![d^D\hat{n}]
  \!\int_{0}^{\infty}\!d\chi\;
  \sinh^D\chi \,
  f\!\left(-i \sinh\chi\, \hat n,\;\cosh\chi\right).
\end{align}
\end{subequations}
Thus we can rewrite
\begin{align}
    I&=I_A+I_B+I_C\nn
    &=i\qty(i^{-D-2}\int_{\EAdS^-} [d^{D+1}y]+ \int_{\dS} [d^{D+1}y]  +i^{-D-2}\int_{\EAdS^+} [d^{D+1}y] ) f(x)
    \label{formula_deform}
\end{align}
where
\begin{align}
  x&= \left\{
  \begin{array}{ll}
  -i y & ( y \in \EAdS), \\
   y   & ( y \in \dS) .
  \end{array}
  \right.
\end{align}

\section{Minkowski Slicing and Contour Deformation}\label{Slicing}
In this section, we consider the integral
\begin{align}
\label{eq:J-def}
J=\int_{\mathbb{R}^{D+2}} [d^{D+2} X]\, f(X).
\end{align}
In polar coordinates, this can be rewritten as
\begin{align}
\label{eq:polar}
    J=\int_{0}^{\infty}\!dr \, r^{D+1}\int_{S^{D+1}} [d^{D+1} x]\, f(rx)\,.
\end{align}
We would like to evaluate \eqref{eq:J-def} in two ways (following appendix \ref{Mink_Slice} and \ref{Contour_deformation}) and verify that, provided $f$ has suitable analyticity in the complex $r$-plane, both methods yield the same result.

\subsection{Minkowski Slicing}\label{Mink_Slice}
We perform the Wick rotation
\begin{align}
\label{ac:time}
  X^{D+2}\;=\; iX^{0}\,.
\end{align}
With this analytic continuation to Minkowski spacetime, we consider the same integral and partition the integration domain into three regions $A,B,C$. Region $B$ is defined by $(X)^{2}>0$, region $A$ by $(X)^{2}<0$ with $X^{0}<0$, and region $C$ by $(X)^{2}<0$ with $X^{0}>0$. Each region can be sliced by either EAdS or dS slices as illustrated in Figure~\ref{Milne}.

\paragraph{Spherical Rindler segment \(B\).}
The contribution from region $B$ is
\begin{align}
  J_{B}\;&=\;
   i \int_{M_{D+2},\;(X)^{2}>0} [d^{ D+2}X]\,  f\!\left(\bm{X},\,iX^{0}\right) \nonumber\\
  &=   i
  \!\int_{0}^{\infty}\!dr \,r^{D+1}
  \!\int_{\dS} [d^{D+1}x]\,
  f\!\left(r\cosh t\, \hat n,\; i r\sinh t \right),
  \label{eq:IC-final}
\end{align}
where we choose coordinates
\(
  X^{0}=r\sinh t,\;
  \bm{X}=r\cosh t\,\hat n
\).

\paragraph{Milne segments \(A\) and \(C\).}
The contributions from these two regions are
\begin{align}
  J_{A}&\;=\;
  i \int_{M_{D+2},\;(X)^{2}<0,\;X^{0}<0} [d^{ D+2}X]\, f\!\left(\bm{X},\,iX^{0}\right)\nonumber\\
  &=i
  \!\int_{0}^{\infty}\!dr \,r^{D+1}
  \!\int_{\text{EAdS}^-} [d^{D+1}x]\,
  f\!\left(r\sinh\chi\, \hat n,\; -i r\cosh\chi\right),
\end{align}
and
\begin{align}
  J_{C}&\;=\;
   i \int_{M_{D+2},\;(X)^{2}<0,\;X^{0}>0} [d^{ D+2}X]\, f\!\left(\bm{X},\,iX^{0}\right)\nonumber\\
   &=i
  \!\int_{0}^{\infty}\!dr \,r^{D+1}
  \!\int_{\text{EAdS}^+} [d^{D+1}x]\,
  f\!\left(r\sinh\chi \,\hat n,\; i r\cosh\chi\right).
  \label{eq:IAE-final}
\end{align}
Region \(A\) is described by the Milne coordinates
\(
   X^{0}=-r\cosh\chi,\;
   X^{i}=r\sinh\chi\,\hat n^{i}
\).
Similarly, region \(C\) is given by
\(
   X^{0}=r\cosh\chi,\;
   X^{i}=r\sinh\chi\,\hat n^{i}
\),
with the same ranges for \(r,\chi,\hat n\). Hence, we finally obtain
\begin{align}
    J&=J_A+J_B+J_C\nn
    &=i\Bigg[\int_{0}^{\infty}\! dr\, r^{D+1}
       \Bigg(\int_{\EAdS^-} [d^{D+1}x]+\int_{\dS} [d^{D+1}x]+\int_{\EAdS^+} [d^{D+1}x]\Bigg)\Bigg]\,f(rx)\,.
    \label{Slicing_master}
\end{align}

\subsection{Contour Deformation}\label{Contour_deformation}
First, deforming the $\theta$-contour of Eq.\eqref{eq:polar} in the complex plane as detailed in Appendix~\ref{app:contour-split}, the angular integral domain splits into $\EAdS^\pm$ and $\dS$ as Eq.\eqref{formula_deform}. Concretely,
\begin{align}
\label{eq:contour-master}
J \;=\; i \int_{0}^{\infty}\!dr\, r^{D+1}
\left(i^{-D-2}\int_{\EAdS^-} [d^{D+1}y]+\int_{\dS} [d^{D+1}y]+i^{-D-2}\int_{\EAdS^+} [d^{D+1}y]\right) f(rx)\,,
\end{align}
From this representation, in particular the $A$ and $C$ contributions can be written as
\begin{subequations}
\begin{align}
 J_{A} &=
  i^{-D-1}
  \!\int_{0}^{\infty}\!dr \, r^{D+1} \int_{\EAdS^-} [d^{D+1}y]\, f(-iry), \\
 J_{C} &=
  i^{-D-1}
  \!\int_{0}^{\infty}\!dr \, r^{D+1} \int_{\EAdS^+} [d^{D+1}y]\, f(-iry),
\end{align}
\end{subequations}
and after the Wick rotation $r\to i r$ in both cases we obtain
\begin{subequations}
\label{I_ABC1}
\begin{align}
  J_{A} &=
  i
  \!\int_{0}^{\infty}\!dr \, r^{D+1}
  \!\int_{\text{EAdS}^-} [d^{D+1}y]\,
  f\!\left(ry\right),\\
  J_{C} &=
  i
  \!\int_{0}^{\infty}\!dr \, r^{D+1}
  \!\int_{\text{EAdS}^+} [d^{D+1} y]\,
  f\!\left(ry\right).
\end{align}
\end{subequations}

Thus, Eq.\eqref{eq:contour-master} eventually coincides with Eq.\eqref{Slicing_master}. Therefore, the evaluation by contour-deforming the angular variable \(\theta\) and the radius \(r\) in the complex plane agrees with the evaluation obtained by first Wick rotating \(X^{D+2}=iX^{0}\), mapping to Minkowski space, and splitting the domain into $A,B,C$. The crucial point is that we performed a Wick rotation in $r$, hence we assume analyticity of $f$ in $r$. In our applications, the integrand contains factors such as $(-2rp\cdot x)^{-\Delta}$, which introduce branch points/cuts and can violate this assumption. Consequently, the two evaluations do not necessarily coincide in general.

\section{Early/Late Correlators in dS}
\label{pro_of_S}

We begin with the Euclidean Green's function of a massive scalar field on
\(\mathrm{S}^{D+1}\) (see, e.g., \cite{Marolf:2010zp}):
\begin{align}
G(x_1,x_2)
   &=
   \frac{\Gamma(\Delta_{+}) \Gamma(\Delta_{-})}
        {2 (2\pi)^{\frac{D+1}{2}} 
         \Gamma\!\left(\frac{D+1}{2}\right)}
   (1-x_1\!\cdot\!x_2)^{\frac{1-D}{2}} 
   {}_2F_{1}\!\left(
      \frac{D+1}{2}-\Delta_{+},
      \frac{D+1}{2}-\Delta_{-};
      \frac{D+1}{2};
      \frac{1+x_1\!\cdot\!x_2}{2}\right),
\label{eq:Sprop_original}
\end{align}
where $x_i$ are the embedding coordinates of $\mathrm{S}^{D+1}$ and \(\Delta_{\pm}\) are
\begin{equation}
\Delta_{\pm}\;=\;\frac{D}{2}\pm i\sqrt{m^{2}-\frac{D^{2}}{4}}.
\label{eq:Delta_pm}
\end{equation}
\eqref{eq:Sprop_original} is also written as
\begin{align}
G(x_1,x_2)
   &=
   2^{\frac{1-D}{2}} 
   \frac{\Gamma(\Delta_{+}) \Gamma(\Delta_{-})}
        {2 (2\pi)^{\frac{D+1}{2}} 
         \Gamma\!\left(\frac{D+1}{2}\right)}
   {}_2F_{1}\!\left(
      \Delta_{+}, 
      \Delta_{-}; 
      \frac{D+1}{2}; 
      \frac{1+x_1\!\cdot\!x_2}{2}\right).
\label{eq:Sprop_Gauss}
\end{align}

We can obtain the two-point function \cite{Bunch:1978yq, Mottola:1984ar, Allen:1985ux} on dS with the Euclidean vacuum, by continuing the polar angle as
\(\theta_i = \frac{\pi}{2} + i t_i\), as
\begin{align}
G(x_1,x_2)
   &=
   \frac{\Gamma(\Delta_{+}) \Gamma(\Delta_{-})}
        {(4\pi)^{\frac{D+1}{2}} 
         \Gamma\!\left(\frac{D+1}{2}\right)}
   {}_2F_{1}\!\left(
      \Delta_{+}, 
      \Delta_{-}; 
      \frac{D+1}{2}; 
      \frac{1+x_1\!\cdot\!x_2-i\epsilon}{2}\right),
\label{eq:dSprop}
\end{align}
where we choose the branch corresponding to the Feynman propagator.

From the free action \eqref{LofS}, we should have
\begin{align}
    \langle \psi_{\lambda_1}^\dagger(x_1) \psi_{\lambda_2}(x_2) \rangle =\delta(\lambda_1-\lambda_2)G(x_1,x_2)
    \label{delta-G}
\end{align}
in dS.

\paragraph{Late-time asymptotics.}
In section \ref{sec4}, we parametrize the embedding coordinates of dS as $x(t,\bm{p})=(\sinh t, \cosh t \bm{p})$ and denote $p=(1, \bm{p})$, $p_-=(1, -\bm{p})$.
Setting \(x_1\simeq \frac12 e^{t_1}p_1\), we obtain, for \(t_1\to+\infty\),
\begin{align}
G(x_1,x_2)\;\sim\;
\frac{1}{4\pi^{h+1}}
\left[
   e^{-\Delta_{+}t_1} 
   2^{\Delta_{+}} 
   \Gamma(h-\Delta_{+}) 
   \Gamma(\Delta_{+}) 
   (-2p_1\!\cdot\!x_2+i\epsilon)^{-\Delta_{+}}
   +(\Delta_{+}\!\leftrightarrow\!\Delta_{-})
\right].
\label{eq:asymptotic_t1}
\end{align}

\paragraph{Integral symmetries.}
We introduce
\begin{equation}
M_{\lambda}
   \;=\;
   2^{-2i\lambda} 
   \frac{\Gamma(h-i\lambda)}{\Gamma(-i\lambda)} 
   i^{2i\lambda} 
   \pi^{-h}.
\label{eq:Mlambda}
\end{equation}
Eq.~\eqref{eq:asymptotic_t1} 
is invariant under the simultaneous transformations 
\begin{align}
F_{+}(p_1)
   &\longrightarrow
   M_{\lambda}\!\int\!D^{d}p' 
   (-2p_1\!\cdot\!p')^{-\Delta_{-}} F_{+}(p')
   \;=\;F_{-}(p_1),
&
e^{-\Delta_{+}t_1}
   &\longrightarrow
   e^{-\Delta_{-}t_1},
\label{eq:symmetry1}\\
F_{-}(p_1)
   &\longrightarrow
   M_{-\lambda}\!\int\!D^{d}p' 
   (-2p_1\!\cdot\!p')^{-\Delta_{+}} F_{-}(p')
   \;=\;F_{+}(p_1),
&
e^{-\Delta_{-}t_1}
   &\longrightarrow
   e^{-\Delta_{+}t_1},
\label{eq:symmetry2}
\end{align}
where $F_\pm$ is the term proportial to $e^{-\Delta_{\pm} t_1}$. For example, in this case,
\begin{equation}
F_{\pm}(p_1)\;=\;
\frac{2^{\Delta_\pm}}{4\pi^{h+1}} 
\Gamma(h-\Delta_{\pm}) 
\Gamma(\Delta_{\pm}) 
(-2p_{1}\!\cdot\!x_2+i\epsilon)^{-\Delta_{\pm}}.
\label{eq:Fpm}
\end{equation}

\paragraph{Late–early correlator and contact terms.}
For \(t_1\to+\infty\) and \(t_2\to-\infty\) with
$x_2\simeq -\frac12 e^{-t_2}p_{2-}$, a naive evaluation gives
\begin{align}
G_{\text{naive}}(x_1,x_2)
   &\sim
   e^{-\Delta_{+}t_1+\Delta_{+}t_2} 
   C_{\Delta_{+}} 
   i^{-2\Delta_{+}} 
   (-2p_1\!\cdot\!p_{2-})^{-\Delta_{+}}
   +(\Delta_{+}\!\leftrightarrow\!\Delta_{-}),
\label{eq:naive}
\end{align}
with
\begin{equation}
C_{\Delta}
   \;=\;
   \frac{2^{2\Delta}}{4\pi^{h+1}} 
   \Gamma(h-\Delta) \Gamma(\Delta),
\label{eq:Cdelta}
\end{equation}
where we used \(\left(-x+i\epsilon\right)^{a}=i^{ 2a}x^{a}\;(x>0)\). The naive equation~\eqref{eq:naive} violates the above symmetries
\eqref{eq:symmetry1}–\eqref{eq:symmetry2}. This is because we implicitly
assumed \(x_1 \cdot x_2 \neq 0\). 
We should consider contributions for the $x_1 \cdot x_2 \sim 0$ case, as we have two contributions from $\bm{p}\sim \bm{k}$ and $\bm{p}\neq \bm{k}$ regions in sec.~\ref{sec4}.
We here evaluate this additional contribution just by restoring the invariance of symmetry. The symmetry indeed requires adding contact terms, giving
\begin{align}
G(x_1,x_2)
   &\sim
   \sum_{\sigma=\pm}
   e^{-\Delta_{\sigma}(t_1+t_2)} 
   C_{\Delta_{\sigma}} 
   i^{-2\Delta_{\sigma}} 
   (-2p_1\!\cdot\!p_{2-})^{-\Delta_{\sigma}}
\nonumber\\
   &\hspace{1.5em}
   +\sum_{\sigma=\pm}
   e^{-\Delta_{\sigma}t_1-\Delta_{-\sigma}t_2} 
   D_{\lambda} 
   i^{-D} 
   \delta^{(D)}(p_1-p_{2-}),
\label{eq:Green_limit}
\end{align}
where
\begin{equation}
D_{\lambda}
   \;=\;
   \frac{2^{D}}{4\pi} 
   {\Gamma(i\lambda) \Gamma(-i\lambda)}.
\label{eq:Dlambda}
\end{equation}
It has been shown that a similar delta-function-like term appears in the two-point function at late times \cite{SalehiVaziri:2024joi,Sengor:2021zlc}.

\paragraph{Boundary two-point functions.}
Multiplying \eqref{eq:Green_limit} by \(\delta(\lambda_1-\lambda_2)\) as \eqref{delta-G} yields the late/early cosmological correlators for a complex scalar field $\psi_\lambda$ with continuous parameter $\lambda$ in \eqref{LofS}:
\begin{align}
\left\langle
  \{{O}^{+}_{\Delta_{1+}}(p_1)\}^\dagger {O}^{-}_{\Delta_{2-}}(p_2)
\right\rangle
   &=
   \delta(\lambda_1-\lambda_2) 
   i^{-2\Delta_{1-}} 
   C_{\Delta_{1-}} 
   (-2p_1\!\cdot\!p_{2})^{-\Delta_{1-}},
\\
\left\langle
  \{{O}^{+}_{\Delta_{1+}}(p_1)\}^\dagger 
  {O}^{-}_{\Delta_{2+}}(p_2)
\right\rangle
   &=
   \delta(\lambda_1-\lambda_2) 
   i^{-D} 
   D_{\lambda_1} 
   \delta^{(D)}(p_1-p_{2}).
\label{eq:boundary_2pt}
\end{align}

\bibliographystyle{utphys} 
\bibliography{ref} 

@article{Cotler:2023qwh,
    author = "Cotler, Jordan and Miller, Noah and Strominger, Andrew",
    title = "{An integer basis for celestial amplitudes}",
    eprint = "2302.04905",
    archivePrefix = "arXiv",
    primaryClass = "hep-th",
    doi = "10.1007/JHEP08(2023)192",
    journal = "JHEP",
    volume = "08",
    pages = "192",
    year = "2023"
}

@article{Freidel:2022skz,
    author = "Freidel, Laurent and Pranzetti, Daniele and Raclariu, Ana-Maria",
    title = "{A discrete basis for celestial holography}",
    eprint = "2212.12469",
    archivePrefix = "arXiv",
    primaryClass = "hep-th",
    doi = "10.1007/JHEP02(2024)176",
    journal = "JHEP",
    volume = "02",
    pages = "176",
    year = "2024"
}

@article{Duary:2022onm,
    author = "Duary, Sarthak",
    title = "{Celestial amplitude for 2d theory}",
    eprint = "2209.02776",
    archivePrefix = "arXiv",
    primaryClass = "hep-th",
    doi = "10.1007/JHEP12(2022)060",
    journal = "JHEP",
    volume = "12",
    pages = "060",
    year = "2022"
}

@article{Jorstad:2023ajr,
    author = "J{\o}rstad, Eivind and Pasterski, Sabrina and Sharma, Atul",
    title = "{Equating extrapolate dictionaries for massless scattering}",
    eprint = "2310.02186",
    archivePrefix = "arXiv",
    primaryClass = "hep-th",
    doi = "10.1007/JHEP02(2024)228",
    journal = "JHEP",
    volume = "02",
    pages = "228",
    year = "2024"
}

@article{Giddings:1999jq,
    author = "Giddings, Steven B.",
    title = "{Flat space scattering and bulk locality in the AdS / CFT correspondence}",
    eprint = "hep-th/9907129",
    archivePrefix = "arXiv",
    doi = "10.1103/PhysRevD.61.106008",
    journal = "Phys. Rev. D",
    volume = "61",
    pages = "106008",
    year = "2000"
}

@article{Susskind:1998vk,
    author = "Susskind, Leonard",
    editor = "Burgess, C. P. and Myers, Robert C.",
    title = "{Holography in the flat space limit}",
    eprint = "hep-th/9901079",
    archivePrefix = "arXiv",
    doi = "10.1063/1.1301570",
    journal = "AIP Conf. Proc.",
    volume = "493",
    number = "1",
    pages = "98--112",
    year = "1999"
}

@article{Polchinski:1999ry,
    author = "Polchinski, Joseph",
    title = "{S matrices from AdS space-time}",
    eprint = "hep-th/9901076",
    archivePrefix = "arXiv",
    reportNumber = "NST-ITP-99-02",
    month = "1",
    year = "1999"
}

@article{Iacobacci:2022yjo,
    author = "Iacobacci, Lorenzo and Sleight, Charlotte and Taronna, Massimo",
    title = "{From celestial correlators to AdS, and back}",
    eprint = "2208.01629",
    archivePrefix = "arXiv",
    primaryClass = "hep-th",
    doi = "10.1007/JHEP06(2023)053",
    journal = "JHEP",
    volume = "06",
    pages = "053",
    year = "2023"
}

@article{Benincasa:2022gtd,
    author = "Benincasa, Paolo",
    title = "{Amplitudes meet Cosmology: A (Scalar) Primer}",
    eprint = "2203.15330",
    archivePrefix = "arXiv",
    primaryClass = "hep-th",
    doi = "10.1142/S0217751X22300101",
    month = "3",
    year = "2022"
}

@article{Loparco:2023rug,
    author = "Loparco, Manuel and Penedones, Joao and Salehi Vaziri, Kamran and Sun, Zimo",
    title = {{The K{\"a}ll{\'e}n-Lehmann representation in de Sitter spacetime}},
    eprint = "2306.00090",
    archivePrefix = "arXiv",
    primaryClass = "hep-th",
    doi = "10.1007/JHEP12(2023)159",
    journal = "JHEP",
    volume = "12",
    pages = "159",
    year = "2023"
}

@article{Hogervorst:2021uvp,
    author = "Hogervorst, Matthijs and Penedones, Jo{\~a}o and Vaziri, Kamran Salehi",
    title = "{Towards the non-perturbative cosmological bootstrap}",
    eprint = "2107.13871",
    archivePrefix = "arXiv",
    primaryClass = "hep-th",
    doi = "10.1007/JHEP02(2023)162",
    journal = "JHEP",
    volume = "02",
    pages = "162",
    year = "2023"
}

@article{Mottola:1984ar,
    author = "Mottola, E.",
    title = "{Particle Creation in de Sitter Space}",
    reportNumber = "NSF-ITP-84-123",
    doi = "10.1103/PhysRevD.31.754",
    journal = "Phys. Rev. D",
    volume = "31",
    pages = "754",
    year = "1985"
}

@article{Allen:1985ux,
    author = "Allen, Bruce",
    title = "{Vacuum States in de Sitter Space}",
    reportNumber = "UCSB-TH-3-1985",
    doi = "10.1103/PhysRevD.32.3136",
    journal = "Phys. Rev. D",
    volume = "32",
    pages = "3136",
    year = "1985"
}

@article{Marolf:2010zp,
    author = "Marolf, Donald and Morrison, Ian A.",
    title = "{The IR stability of de Sitter: Loop corrections to scalar propagators}",
    eprint = "1006.0035",
    archivePrefix = "arXiv",
    primaryClass = "gr-qc",
    doi = "10.1103/PhysRevD.82.105032",
    journal = "Phys. Rev. D",
    volume = "82",
    pages = "105032",
    year = "2010"
}

@article{Fukuma:2013mx,
    author = "Fukuma, Masafumi and Sugishita, Sotaro and Sakatani, Yuho",
    title = "{Propagators in de Sitter space}",
    eprint = "1301.7352",
    archivePrefix = "arXiv",
    primaryClass = "hep-th",
    reportNumber = "KUNS-2432, MISC-2013-01",
    doi = "10.1103/PhysRevD.88.024041",
    journal = "Phys. Rev. D",
    volume = "88",
    number = "2",
    pages = "024041",
    year = "2013"
}

@article{Sleight:2020obc,
    author = "Sleight, Charlotte and Taronna, Massimo",
    title = "{From AdS to dS exchanges: Spectral representation, Mellin amplitudes, and crossing}",
    eprint = "2007.09993",
    archivePrefix = "arXiv",
    primaryClass = "hep-th",
    doi = "10.1103/PhysRevD.104.L081902",
    journal = "Phys. Rev. D",
    volume = "104",
    number = "8",
    pages = "L081902",
    year = "2021"
}

@article{Hirai:2018ijc,
    author = "Hirai, Hayato and Sugishita, Sotaro",
    title = "{Conservation Laws from Asymptotic Symmetry and Subleading Charges in QED}",
    eprint = "1805.05651",
    archivePrefix = "arXiv",
    primaryClass = "hep-th",
    reportNumber = "OU-HET-971",
    doi = "10.1007/JHEP07(2018)122",
    journal = "JHEP",
    volume = "07",
    pages = "122",
    year = "2018"
}

@article{Campiglia:2015qka,
    author = "Campiglia, Miguel and Laddha, Alok",
    title = "{Asymptotic symmetries of QED and Weinberg{\textquoteright}s soft photon theorem}",
    eprint = "1505.05346",
    archivePrefix = "arXiv",
    primaryClass = "hep-th",
    doi = "10.1007/JHEP07(2015)115",
    journal = "JHEP",
    volume = "07",
    pages = "115",
    year = "2015"
}

@article{DiPietro:2021sjt,
    author = "Di Pietro, Lorenzo and Gorbenko, Victor and Komatsu, Shota",
    title = "{Analyticity and unitarity for cosmological correlators}",
    eprint = "2108.01695",
    archivePrefix = "arXiv",
    primaryClass = "hep-th",
    reportNumber = "CERN-TH-2021-118",
    doi = "10.1007/JHEP03(2022)023",
    journal = "JHEP",
    volume = "03",
    pages = "023",
    year = "2022"
}

@article{Dey:2024zjx,
    author = "Dey, Indranil and Nanda, Kanhu Kishore and Roy, Akashdeep and Trivedi, Sandip P.",
    title = "{Aspects of dS/CFT holography}",
    eprint = "2407.02417",
    archivePrefix = "arXiv",
    primaryClass = "hep-th",
    doi = "10.1007/JHEP05(2025)168",
    journal = "JHEP",
    volume = "05",
    pages = "168",
    year = "2025"
}

@article{Isono:2020qew,
    author = "Isono, Hiroshi and Liu, Hoiki Madison and Noumi, Toshifumi",
    title = "{Wavefunctions in dS/CFT revisited: principal series and double-trace deformations}",
    eprint = "2011.09479",
    archivePrefix = "arXiv",
    primaryClass = "hep-th",
    reportNumber = "UT-Komaba/20-4, KOBE-COSMO-20-11",
    doi = "10.1007/JHEP04(2021)166",
    journal = "JHEP",
    volume = "04",
    pages = "166",
    year = "2021"
}

@article{Sleight:2021plv,
    author = "Sleight, Charlotte and Taronna, Massimo",
    title = "{From dS to AdS and back}",
    eprint = "2109.02725",
    archivePrefix = "arXiv",
    primaryClass = "hep-th",
    doi = "10.1007/JHEP12(2021)074",
    journal = "JHEP",
    volume = "12",
    pages = "074",
    year = "2021"
}

@article{Higuchi:2010xt,
    author = "Higuchi, Atsushi and Marolf, Donald and Morrison, Ian A.",
    title = "{On the Equivalence between Euclidean and In-In Formalisms in de Sitter QFT}",
    eprint = "1012.3415",
    archivePrefix = "arXiv",
    primaryClass = "gr-qc",
    doi = "10.1103/PhysRevD.83.084029",
    journal = "Phys. Rev. D",
    volume = "83",
    pages = "084029",
    year = "2011"
}

@article{SalehiVaziri:2024joi,
    author = "Salehi Vaziri, Kamran",
    title = "{A non-perturbative construction of the de Sitter late-time boundary}",
    eprint = "2412.00183",
    archivePrefix = "arXiv",
    primaryClass = "hep-th",
    month = "11",
    year = "2024"
}

@article{tHooft:1993dmi,
    author = "'t Hooft, Gerard",
    title = "{Dimensional reduction in quantum gravity}",
    eprint = "gr-qc/9310026",
    archivePrefix = "arXiv",
    reportNumber = "THU-93-26",
    journal = "Conf. Proc. C",
    volume = "930308",
    pages = "284--296",
    year = "1993"
}

@article{Susskind:1994vu,
    author = "Susskind, Leonard",
    title = "{The World as a hologram}",
    eprint = "hep-th/9409089",
    archivePrefix = "arXiv",
    reportNumber = "SU-ITP-94-33",
    doi = "10.1063/1.531249",
    journal = "J. Math. Phys.",
    volume = "36",
    pages = "6377--6396",
    year = "1995"
}

@article{Maldacena:1997re,
    author = "Maldacena, Juan Martin",
    title = "{The Large $N$ limit of superconformal field theories and supergravity}",
    eprint = "hep-th/9711200",
    archivePrefix = "arXiv",
    reportNumber = "HUTP-97-A097, HUTP-98-A097",
    doi = "10.4310/ATMP.1998.v2.n2.a1",
    journal = "Adv. Theor. Math. Phys.",
    volume = "2",
    pages = "231--252",
    year = "1998"
}

@article{Gubser:1998bc,
    author = "Gubser, S. S. and Klebanov, Igor R. and Polyakov, Alexander M.",
    title = "{Gauge theory correlators from noncritical string theory}",
    eprint = "hep-th/9802109",
    archivePrefix = "arXiv",
    reportNumber = "PUPT-1767",
    doi = "10.1016/S0370-2693(98)00377-3",
    journal = "Phys. Lett. B",
    volume = "428",
    pages = "105--114",
    year = "1998"
}

@article{Witten:1998qj,
    author = "Witten, Edward",
    title = "{Anti de Sitter space and holography}",
    eprint = "hep-th/9802150",
    archivePrefix = "arXiv",
    reportNumber = "IASSNS-HEP-98-15",
    doi = "10.4310/ATMP.1998.v2.n2.a2",
    journal = "Adv. Theor. Math. Phys.",
    volume = "2",
    pages = "253--291",
    year = "1998"
}

@article{Strominger:2001pn,
    author = "Strominger, Andrew",
    title = "{The dS / CFT correspondence}",
    eprint = "hep-th/0106113",
    archivePrefix = "arXiv",
    doi = "10.1088/1126-6708/2001/10/034",
    journal = "JHEP",
    volume = "10",
    pages = "034",
    year = "2001"
}

@article{Maldacena:2002vr,
    author = "Maldacena, Juan Martin",
    title = "{Non-Gaussian features of primordial fluctuations in single field inflationary models}",
    eprint = "astro-ph/0210603",
    archivePrefix = "arXiv",
    doi = "10.1088/1126-6708/2003/05/013",
    journal = "JHEP",
    volume = "05",
    pages = "013",
    year = "2003"
}

@article{Pasterski:2016qvg,
    author = "Pasterski, Sabrina and Shao, Shu-Heng and Strominger, Andrew",
    title = "{Flat Space Amplitudes and Conformal Symmetry of the Celestial Sphere}",
    eprint = "1701.00049",
    archivePrefix = "arXiv",
    primaryClass = "hep-th",
    doi = "10.1103/PhysRevD.96.065026",
    journal = "Phys. Rev. D",
    volume = "96",
    number = "6",
    pages = "065026",
    year = "2017"
}

@book{Strominger:2017zoo,
    author = "Strominger, Andrew",
    title = "{Lectures on the Infrared Structure of Gravity and Gauge Theory}",
    eprint = "1703.05448",
    archivePrefix = "arXiv",
    primaryClass = "hep-th",
    isbn = "978-0-691-17973-5",
    month = "3",
    year = "2017"
}

@article{Pasterski:2017kqt,
    author = "Pasterski, Sabrina and Shao, Shu-Heng",
    title = "{Conformal basis for flat space amplitudes}",
    eprint = "1705.01027",
    archivePrefix = "arXiv",
    primaryClass = "hep-th",
    doi = "10.1103/PhysRevD.96.065022",
    journal = "Phys. Rev. D",
    volume = "96",
    number = "6",
    pages = "065022",
    year = "2017"
}

@article{Pasterski:2017ylz,
    author = "Pasterski, Sabrina and Shao, Shu-Heng and Strominger, Andrew",
    title = "{Gluon Amplitudes as 2d Conformal Correlators}",
    eprint = "1706.03917",
    archivePrefix = "arXiv",
    primaryClass = "hep-th",
    doi = "10.1103/PhysRevD.96.085006",
    journal = "Phys. Rev. D",
    volume = "96",
    number = "8",
    pages = "085006",
    year = "2017"
}

@article{Donnay:2020guq,
    author = "Donnay, Laura and Pasterski, Sabrina and Puhm, Andrea",
    title = "{Asymptotic Symmetries and Celestial CFT}",
    eprint = "2005.08990",
    archivePrefix = "arXiv",
    primaryClass = "hep-th",
    reportNumber = "CPHT-RR022.042020",
    doi = "10.1007/JHEP09(2020)176",
    journal = "JHEP",
    volume = "09",
    pages = "176",
    year = "2020"
}

@article{Ogawa:2022fhy,
    author = "Ogawa, Naoki and Takayanagi, Tadashi and Tsuda, Takashi and Waki, Takahiro",
    title = "{Wedge holography in flat space and celestial holography}",
    eprint = "2207.06735",
    archivePrefix = "arXiv",
    primaryClass = "hep-th",
    reportNumber = "YITP-22-71, IPMU22-0036",
    doi = "10.1103/PhysRevD.107.026001",
    journal = "Phys. Rev. D",
    volume = "107",
    number = "2",
    pages = "026001",
    year = "2023"
}

@article{Furugori:2023hgv,
    author = "Furugori, Hideo and Ogawa, Naoki and Sugishita, Sotaro and Waki, Takahiro",
    title = "{Celestial two-point functions and rectified dictionary}",
    eprint = "2312.07057",
    archivePrefix = "arXiv",
    primaryClass = "hep-th",
    reportNumber = "KUNS-2988, YITP-23-160",
    doi = "10.1007/JHEP02(2024)063",
    journal = "JHEP",
    volume = "02",
    pages = "063",
    year = "2024"
}

@article{Cheung:2016iub,
    author = "Cheung, Clifford and de la Fuente, Anton and Sundrum, Raman",
    title = "{4D scattering amplitudes and asymptotic symmetries from 2D CFT}",
    eprint = "1609.00732",
    archivePrefix = "arXiv",
    primaryClass = "hep-th",
    reportNumber = "CALT-TH-2016-024, UMD-PP-017-010",
    doi = "10.1007/JHEP01(2017)112",
    journal = "JHEP",
    volume = "01",
    pages = "112",
    year = "2017"
}

@article{Banks:1998dd,
    author = "Banks, Tom and Douglas, Michael R. and Horowitz, Gary T. and Martinec, Emil J.",
    title = "{AdS dynamics from conformal field theory}",
    eprint = "hep-th/9808016",
    archivePrefix = "arXiv",
    reportNumber = "NSF-ITP-98-082, EFI-98-30",
    month = "8",
    year = "1998"
}

@article{Bousso:2001mw,
    author = "Bousso, Raphael and Maloney, Alexander and Strominger, Andrew",
    title = "{Conformal vacua and entropy in de Sitter space}",
    eprint = "hep-th/0112218",
    archivePrefix = "arXiv",
    doi = "10.1103/PhysRevD.65.104039",
    journal = "Phys. Rev. D",
    volume = "65",
    pages = "104039",
    year = "2002"
}

@article{Bunch:1978yq,
    author = "Bunch, T. S. and Davies, P. C. W.",
    title = "{Quantum Field Theory in de Sitter Space: Renormalization by Point Splitting}",
    doi = "10.1098/rspa.1978.0060",
    journal = "Proc. Roy. Soc. Lond. A",
    volume = "360",
    pages = "117--134",
    year = "1978"
}

@article{Marolf:2010nz,
    author = "Marolf, Donald and Morrison, Ian A.",
    title = "{The IR stability of de Sitter QFT: results at all orders}",
    eprint = "1010.5327",
    archivePrefix = "arXiv",
    primaryClass = "gr-qc",
    doi = "10.1103/PhysRevD.84.044040",
    journal = "Phys. Rev. D",
    volume = "84",
    pages = "044040",
    year = "2011"
}

@article{deBoer:2003vf,
    author = "de Boer, Jan and Solodukhin, Sergey N.",
    title = "{A Holographic reduction of Minkowski space-time}",
    eprint = "hep-th/0303006",
    archivePrefix = "arXiv",
    reportNumber = "ITFA-2003-11",
    doi = "10.1016/S0550-3213(03)00494-2",
    journal = "Nucl. Phys. B",
    volume = "665",
    pages = "545--593",
    year = "2003"
}

@article{Simmons-Duffin:2012juh,
    author = "Simmons-Duffin, David",
    title = "{Projectors, Shadows, and Conformal Blocks}",
    eprint = "1204.3894",
    archivePrefix = "arXiv",
    primaryClass = "hep-th",
    doi = "10.1007/JHEP04(2014)146",
    journal = "JHEP",
    volume = "04",
    pages = "146",
    year = "2014"
}

@article{Sleight:2023ojm,
    author = "Sleight, Charlotte and Taronna, Massimo",
    title = "{Celestial Holography Revisited}",
    eprint = "2301.01810",
    archivePrefix = "arXiv",
    primaryClass = "hep-th",
    doi = "10.1103/PhysRevLett.133.241601",
    journal = "Phys. Rev. Lett.",
    volume = "133",
    number = "24",
    pages = "241601",
    year = "2024"
}

@article{Iacobacci:2024nhw,
    author = "Iacobacci, Lorenzo and Sleight, Charlotte and Taronna, Massimo",
    title = {{Celestial holography revisited. Part II. Correlators and K\"all\'en-Lehmann}},
    eprint = "2401.16591",
    archivePrefix = "arXiv",
    primaryClass = "hep-th",
    doi = "10.1007/JHEP08(2024)033",
    journal = "JHEP",
    volume = "08",
    pages = "033",
    year = "2024"
}

@article{Kapec:2016jld,
    author = "Kapec, Daniel and Mitra, Prahar and Raclariu, Ana-Maria and Strominger, Andrew",
    title = "{2D Stress Tensor for 4D Gravity}",
    eprint = "1609.00282",
    archivePrefix = "arXiv",
    primaryClass = "hep-th",
    doi = "10.1103/PhysRevLett.119.121601",
    journal = "Phys. Rev. Lett.",
    volume = "119",
    number = "12",
    pages = "121601",
    year = "2017"
}

@article{Balasubramanian:2001nb,
    author = "Balasubramanian, Vijay and de Boer, Jan and Minic, Djordje",
    title = "{Mass, entropy and holography in asymptotically de Sitter spaces}",
    eprint = "hep-th/0110108",
    archivePrefix = "arXiv",
    reportNumber = "VPI-IPPAP-01-01, UPR-964-T",
    doi = "10.1103/PhysRevD.65.123508",
    journal = "Phys. Rev. D",
    volume = "65",
    pages = "123508",
    year = "2002"
}

@article{Henningson:1998gx,
    author = "Henningson, M. and Skenderis, K.",
    title = "{The Holographic Weyl anomaly}",
    eprint = "hep-th/9806087",
    archivePrefix = "arXiv",
    reportNumber = "CERN-TH-98-188, KUL-TF-98-21",
    doi = "10.1088/1126-6708/1998/07/023",
    journal = "JHEP",
    volume = "07",
    pages = "023",
    year = "1998"
}

@article{Balasubramanian:1999re,
    author = "Balasubramanian, Vijay and Kraus, Per",
    title = "{A Stress tensor for Anti-de Sitter gravity}",
    eprint = "hep-th/9902121",
    archivePrefix = "arXiv",
    reportNumber = "HUTP-99-A002, EFI-99-6, NSF-ITP-98-132",
    doi = "10.1007/s002200050764",
    journal = "Commun. Math. Phys.",
    volume = "208",
    pages = "413--428",
    year = "1999"
}

@article{Polyakov:1981rd,
    author = "Polyakov, Alexander M.",
    editor = "Khalatnikov, I. M. and Mineev, V. P.",
    title = "{Quantum Geometry of Bosonic Strings}",
    reportNumber = "Print-81-0351 (LANDAU INST)",
    doi = "10.1016/0370-2693(81)90743-7",
    journal = "Phys. Lett. B",
    volume = "103",
    pages = "207--210",
    year = "1981"
}

@article{Sengor:2021zlc,
    author = "Sengor, Gizem and Skordis, Constantinos",
    title = "{Scalar two-point functions at the late-time boundary of de Sitter}",
    eprint = "2110.01635",
    archivePrefix = "arXiv",
    primaryClass = "hep-th",
    doi = "10.1007/JHEP02(2024)076",
    journal = "JHEP",
    volume = "02",
    pages = "076",
    year = "2024"
}

@article{Brown:1986nw,
    author = "Brown, J. David and Henneaux, M.",
    title = "{Central Charges in the Canonical Realization of Asymptotic Symmetries: An Example from Three-Dimensional Gravity}",
    doi = "10.1007/BF01211590",
    journal = "Commun. Math. Phys.",
    volume = "104",
    pages = "207--226",
    year = "1986"
}

@article{Strominger:1997eq,
    author = "Strominger, Andrew",
    title = "{Black hole entropy from near horizon microstates}",
    eprint = "hep-th/9712251",
    archivePrefix = "arXiv",
    reportNumber = "HUTP-97-A106",
    doi = "10.1088/1126-6708/1998/02/009",
    journal = "JHEP",
    volume = "02",
    pages = "009",
    year = "1998"
}

@article{Harlow:2011ke,
    author = "Harlow, Daniel and Stanford, Douglas",
    title = "{Operator Dictionaries and Wave Functions in AdS/CFT and dS/CFT}",
    eprint = "1104.2621",
    archivePrefix = "arXiv",
    primaryClass = "hep-th",
    reportNumber = "SU-ITP-11-22",
    month = "4",
    year = "2011"
}

@inproceedings{Witten:2001kn,
    author = "Witten, Edward",
    title = "{Quantum gravity in de Sitter space}",
    booktitle = "{Strings 2001: International Conference}",
    eprint = "hep-th/0106109",
    archivePrefix = "arXiv",
    month = "6",
    year = "2001"
}

@article{Goldar:2024crc,
    author = "Goldar, Arundhati and Kajuri, Nirmalya",
    title = "{Bulk reconstruction in de Sitter spacetime}",
    eprint = "2405.16832",
    archivePrefix = "arXiv",
    primaryClass = "hep-th",
    doi = "10.1016/j.physletb.2025.139655",
    journal = "Phys. Lett. B",
    volume = "868",
    pages = "139655",
    year = "2025"
}

@article{Xiao:2014uea,
    author = "Xiao, Xiao",
    title = "{Holographic representation of local operators in de sitter space}",
    eprint = "1402.7080",
    archivePrefix = "arXiv",
    primaryClass = "hep-th",
    doi = "10.1103/PhysRevD.90.024061",
    journal = "Phys. Rev. D",
    volume = "90",
    number = "2",
    pages = "024061",
    year = "2014"
}

@article{Ball:2019atb,
    author = "Ball, Adam and Himwich, Elizabeth and Narayanan, Sruthi A. and Pasterski, Sabrina and Strominger, Andrew",
    title = "{Uplifting AdS$_{3}$/CFT$_{2}$ to flat space holography}",
    eprint = "1905.09809",
    archivePrefix = "arXiv",
    primaryClass = "hep-th",
    doi = "10.1007/JHEP08(2019)168",
    journal = "JHEP",
    volume = "08",
    pages = "168",
    year = "2019"
}

@article{Crawley:2021ivb,
    author = "Crawley, Erin and Miller, Noah and Narayanan, Sruthi A. and Strominger, Andrew",
    title = "{State-operator correspondence in celestial conformal field theory}",
    eprint = "2105.00331",
    archivePrefix = "arXiv",
    primaryClass = "hep-th",
    doi = "10.1007/JHEP09(2021)132",
    journal = "JHEP",
    volume = "09",
    pages = "132",
    year = "2021"
}

\end{document}